\documentclass[12pt]{article}

\usepackage{comment}
\usepackage{tocloft}
\usepackage{graphicx}
\graphicspath{ {./Plots/} }
\usepackage{natbib}
\usepackage[protrusion=true,expansion=true]{microtype}
\usepackage{amsmath,amsthm,amsfonts,amssymb,mathrsfs,dsfont,mathtools}
\usepackage{booktabs}
\usepackage{url}
\usepackage{float}
\usepackage{color}
\usepackage{parskip}
\usepackage[explicit]{titlesec}
\usepackage{threeparttable}
\usepackage{setspace}
\def\doublespacing{\setstretch{2}}
\usepackage[margin=1.00in]{geometry}
\usepackage[font = onehalfspacing]{caption}
\usepackage[usenames,dvipsnames]{xcolor}
\usepackage[hyperfootnotes=true]{hyperref}
\usepackage[labelfont=bf]{caption}
\usepackage{authblk}
\usepackage{caption}
\usepackage[titletoc,toc,title]{appendix}
\usepackage{enumerate}
\usepackage{soul}
\usepackage{subfig}
\usepackage{accents}
\usepackage{subfiles} 
\usepackage[capitalise, nameinlink]{cleveref}
\usepackage{dsfont}
\usepackage{arydshln}

\usepackage{changepage}
\usepackage[noend]{algpseudocode}
\usepackage{tocloft}
\usepackage[protrusion=true,expansion=true]{microtype}
\usepackage{amsmath,amsthm,amsfonts,amssymb,mathrsfs,dsfont,mathtools}
\usepackage{float}
\usepackage{color}
\usepackage{parskip}
\usepackage[explicit]{titlesec}
\usepackage{threeparttable}
\usepackage{setspace}
\usepackage[font = onehalfspacing]{caption}
\usepackage[usenames,dvipsnames]{xcolor}
\usepackage[hyperfootnotes=true]{hyperref}
\usepackage[labelfont=bf]{caption}
\usepackage{authblk}
\usepackage[titletoc,toc,title]{appendix}
\usepackage{enumerate}
\usepackage{soul}
\usepackage{subfig}
\usepackage{accents}
\usepackage[capitalise, nameinlink]{cleveref}
\usepackage{diagbox}
\usepackage{array,multirow}

\usepackage[utf8]{inputenc} % allow utf-8 input
\usepackage[T1]{fontenc}    % use 8-bit T1 fonts
\usepackage{hyperref}       % hyperlinks
\usepackage{url}            % simple URL typesetting
\usepackage{booktabs}       % professional-quality tables
\usepackage{amsfonts}       % blackboard math symbols
\usepackage{nicefrac}       % compact symbols for 1/2, etc.
\usepackage{microtype}      % microtypography
\usepackage{lipsum}		% Can be removed after putting your text content 
\usepackage{graphicx}
\usepackage{natbib}
\usepackage{doi}
\usepackage{subfiles}
\usepackage{bbm}

\usepackage{adjustbox}
\usepackage{multirow}
\usepackage{caption}
\usepackage{amsmath}
\usepackage[section]{placeins}
\usepackage{algorithm}
\usepackage{algpseudocode}
\usepackage{dsfont}

\DeclareUnicodeCharacter{2212}{-}

\newcommand{\bc}[1]{{\color{blue}{#1}}}

\setstcolor{red} 

\hypersetup{colorlinks = true, citecolor = MidnightBlue, linkcolor = MidnightBlue, urlcolor = MidnightBlue}

\titleformat{\section}{\normalfont\large\bfseries}{\thesection}{1em}{#1}
\titleformat{\subsection}{\normalfont\normalsize\bfseries}{\thesubsection}{1em}{#1}
\titleformat{\subsubsection}{\normalfont\normalsize\itshape}{\thesubsubsection}{1em}{#1}
\titlespacing\section{0pt}{12pt plus 4pt minus 2pt}{6pt plus 2pt minus 2pt}
\titlespacing\subsection{0pt}{12pt plus 4pt minus 2pt}{3pt plus 2pt minus 3pt}
\titlespacing\subsubsection{0pt}{12pt plus 4pt minus 2pt}{0pt plus 2pt minus 3pt}

\def\boxit#1{\vbox{\hrule\hbox{\vrule\kern6pt
          \vbox{\kern6pt#1\kern6pt}\kern6pt\vrule}\hrule}}

\definecolor{orange}{rgb}{1,0.5,0}
\definecolor{MyDarkBlue}{rgb}{0,0.08,0.45}

\def\boxit#1{\vbox{\hrule\hbox{\vrule\kern6pt
          \vbox{\kern6pt#1\kern6pt}\kern6pt\vrule}\hrule}}

\definecolor{orange}{rgb}{1,0.5,0}
\definecolor{MyDarkBlue}{rgb}{0,0.08,0.45}

\begin{document}

\title{\vspace{-25pt} \Large \bfseries %Implied-Volatility-Surface-Informed Deep Hedging with Options% 
Deep Hedging with Options \\ Using the Implied Volatility Surface\thanks{François is supported by a fellowship from the Canadian Institute of Derivatives. Gauthier is supported by the Natural Sciences and Engineering Research Council of Canada (NSERC, RGPIN-2024-03791), a professorship funded by HEC Montr\'eal, and the HEC Montr\'eal Foundation. Godin is funded by NSERC (RGPIN-2024-04593).}
 } %This research was enabled in part by support provided by Calcul Québec (https://www.calculquebec.ca/) and Compute Canada (www.computecanada.ca).
\vspace{-5pt}
\author[a]{Pascal Fran\c{c}ois}
\author[b]{Genevi\`eve Gauthier} 
\author[,c]{Fr\'ed\'eric Godin\thanks{Corresponding author. Email:  \href{mailto:frederic.godin@concordia.ca}{frederic.godin@concordia.ca}.}}  
\author[c]{Carlos Octavio P\'erez-Mendoza}  

\renewcommand\Authands{\\} % Ensures a line break between author groups

%\author[a,b,c]{MANYPEOPLE\thanks{Corresponding author. \vspace{0.2em} \newline
%{\mbox{\hspace{0.47cm}} \it Email addresses:} 
%ADD EVERYBODY, \href{mailto:frederic.godin@concordia.ca}{frederic.godin@concordia.ca} (Fr\'ed\'eric Godin).  }}

%\affil[a]{{\small University of Toronto, Rotman School of Management, Toronto, Canada}}
\affil[a]{{\small Department of Finance, HEC Montr\'eal, Montreal, Canada}}
\affil[b]{{\small GERAD and Department of Decision Sciences, HEC Montr\'eal, Montreal, Canada}}
\affil[c]{{\small Concordia University, Department of Mathematics and Statistics, Montreal, Canada}}
%\affil[d]{{\small Quantact Laboratory, Centre de Recherches Math\'ematiques, Montreal, Canada}}

\vspace{-10pt}
\date{ %\bigskip %\bigskip\bigskip\bigskip \bigskip 
\today}

%%%%%%%%%%%%%%%%%%%%%%%%%%%%%%%%%%%%%%%%%%%%%%%%%%%%%%%%%%%%%%%%%%
%%%%%%%%%%%%%%%%%%%%%%%%%%%%%%%%%%%%%%%%%%%%%%%%%%%%%%%%%%%%%%%%%%
%%% ABSTRACT PAGE
%%%%%%%%%%%%%%%%%%%%%%%%%%%%%%%%%%%%%%%%%%%%%%%%%%%%%%%%%%%%%%%%%%
%%%%%%%%%%%%%%%%%%%%%%%%%%%%%%%%%%%%%%%%%%%%%%%%%%%%%%%%%%%%%%%%%%

\maketitle \thispagestyle{empty} %\vfill \pagebreak \vspace*{\fill}

%\begin{center} \Large \bfseries Local hedging of variable annuities in the \\ presence of basis risk  \end{center}
%
%\bigskip \bigskip \bigskip \bigskip

%\vspace{-15pt}
\vspace{-10pt}
\begin{abstract}
\vspace{-5pt}

We propose a deep hedging framework for index option portfolios, grounded in a realistic market simulator that captures the joint dynamics of S\&P 500 returns and the full implied volatility surface. Our approach integrates surface-informed decisions with multiple hedging instruments and explicitly accounts for transaction costs. The hedging strategy also considers the variance risk premium embedded in the hedging instruments, enabling more informed and adaptive risk management. Tested on a historical out-of-sample set of straddles from 2020 to 2023, our method consistently outperforms traditional delta-gamma hedging strategies across a range of market conditions. 

\bigskip %\bigskip %\bigskip

%\noindent \textbf{JEL classification:} E43, G12.
\vspace{-5pt}
%\vspace{0.08in}

\noindent \textbf{JEL classification:} C45, C61, G32.

%\vspace{0.08in}

\noindent \textbf{Keywords:} Deep reinforcement learning, optimal hedging, implied volatility surfaces.

\end{abstract}

\medskip

\thispagestyle{empty} \vfill \pagebreak

\setcounter{page}{1}
\pagenumbering{roman}

\doublespacing

\setcounter{page}{1}
\pagenumbering{arabic}

%%%%%%%%%%%%%%%%%%%%%%%%%%%%%%%%%%%%%%%%%%%%%%%%%%%%%%%%%%%%%%%%%%%%%%%%%
%%%%%%%%%%%%%%%%%%%%%%%%%%%%%%%%%%%%%%%%%%%%%%%%%%%%%%%%%%%%%%%%%%%%%%%%%
%%%%%%%%%%%%%%%%%%%%%%%%%%%%%%%%%%%%%%%%%%%%%%%%%%%%%%%%%%%%%%%%%%%%%%%%%
%%%%%%%%%%%%%%%%%%%%%%%%%%%%%%%%%%%%%%%%%%%%%%%%%%%%%%%%%%%%%%%%%%%%%%%%%
\doublespacing%%%activate before the Introduction and after the abstract

\section{Introduction}\label{se:intro}

%%%%%%%%%%%%%%%%%%%%%%
Hedging decisions are inherently tied to the information available at the time they are made. Traditional approaches typically rely on dynamics of the underlying asset, which are estimated on historical data. Some studies extend this framework by incorporating localized information from the implied volatility (IV) surface, such as at-the-money or short-term implied volatilities \citep{BATES2005195, ALEXANDER2007, franccois2021smile}. In this paper, we address the hedging problem for an index option portfolio using a richer set of information—namely, characteristics of the full implied volatility surface. By exploiting the structure of the entire surface, we aim to better capture market expectations and variance dynamics. 

Capturing such information at each decision point increases the dimensionality of the state vector. This makes reinforcement learning (RL) a natural choice for identifying optimal hedging strategies.
Deep hedging, introduced by \cite{buehler2019deep}, leverages deep reinforcement learning to dynamically adapt to evolving market conditions, capturing both shifting expectations and historical patterns. While this approach has shown remarkable flexibility and adaptability (e.g., \cite{du2020deep}, \cite{cao2020deep}, \cite{carbonneau2021deep}, \cite{wu2023robust}, \cite{cao2023gamma}), the training of the neural network requires a market simulator. \cite{François2025} demonstrate that deep hedging strategies can effectively mitigate transaction costs while incorporating information from the implied volatility (IV) surface. Their study, however, focuses on the hedging of European options using only the underlying asset. The potential benefits of expanding the hedging set to include additional instruments, alongside IV-informed policies, remain unexplored.

We build on the framework introduced by \cite{François2025}, extending it to address the risk management of index option portfolios through the inclusion of an additional hedging instrument.
%\fredcomment{It seems we've already discuss hedging instruments choices above, and go back to it. We could improve the flow of the intro.}
This extension introduces significant challenges, both computational and conceptual. First, the state vector requires further information about the additional hedging instrument and the portfolio to be hedged. 
%The high-dimensional nature of the problem is addressed using risk-aware reinforcement learning, which is well-suited for sequential decision-making under uncertainty.
Second, to ensure that the RL agent learns a true hedging strategy rather than engaging in speculative behavior, we introduce penalty terms in the reward function that discourages excessive risk-taking. This design helps steer the agent toward strategies that align with the core objective of minimizing portfolio risk in a realistic trading environment.

Our study is distinctive in that it simultaneously leverages rich information derived from the implied volatility surface and its dynamics, explicitly accounts for transaction costs—which are particularly significant when trading options—and departs from traditional portfolio tracking by adopting a global hedging objective focused on minimizing terminal hedging error.

Numerical results for hedging a short position on a straddle show that all our RL algorithms consistently and substantially outperform the practitioner's delta and delta-gamma approaches.
In some cases, the RL agent relying only on the underlying as the hedging instrument even outperforms delta-gamma hedging: this happens in particular in the presence of transaction costs when a tail risk performance metric is considered. 

The outperformance of RL approaches can be attributed to several factors. First, RL strategies typically rely on smaller trades. This more gradual rebalancing reduces the likelihood of having to unwind large positions shortly after they are established. Second, the early-stage divergence between RL and delta-gamma positions reflects the RL agent’s deliberate efforts to limit short exposure to the variance risk premium embedded in the option used for hedging.
Thanks to our enriched informational state vector, the RL agent learns and adapts to the time-varying variance risk premium, which is a key driver of hedging costs. 
The impact of risk premia materializes over the long term and is therefore not captured by myopic Greeks-based approaches.

As the model is trained on market data from 1996 to 2020, we use recent option data from 2021 to 2023 to evaluate whether our trained RL algorithm maintains its performance out-of-sample. For this backtesting study, we introduce a new benchmark: the RL algorithm without IV information. We demonstrate the superiority of the RL algorithms with the full information over both the practitioners’ delta-gamma strategy and the RL algorithms with limited information. The RL algorithms without IV information do not outperform the practitioners’ delta-gamma approach in terms of mean squared hedging errors. These results highlight the importance of feeding relevant market information to the RL hedging agent.

The paper is organized as follows. \cref{se:hedgepro} frames the hedging problem in terms of a deep reinforcement learning framework. \cref{se:JMD} provides the components of the market simulator. \cref{se:SSG} presents the numerical results.\footnote{The Python code to replicate the numerical experiments from this paper can be found at the following link: \mbox{\href{https://github.com/cpmendoza/deep-hedging_with_options.git}{https://github.com/cpmendoza/deep-hedging\_with\_options.git}.}} \cref{subsub:benchmarking_realpaths} presents the out-of-sample backtesting results. \cref{se:conclusion} concludes.

%%%%%%%%%%%%%%%%%%%%%%%%%%%%%%%%%%%%%%%%%%%%%%%%%%%%%%%%%%%%%%%%%%%%%%%%%
%%%%%%%%%%%%%%%%%%%%%%%%%%%%%%%%%%%%%%%%%%%%%%%%%%%%%%%%%%%%%%%%%%%%%%%%%
%%%%%%%%%%%%%%%%%%%%%%%%%%%%%%%%%%%%%%%%%%%%%%%%%%%%%%%%%%%%%%%%%%%%%%%%%
%%%%%%%%%%%%%%%%%%%%%%%%%%%%%%%%%%%%%%%%%%%%%%%%%%%%%%%%%%%%%%%%%%%%%%%%%

\section{Deep hedging framework}\label{se:hedgepro}

In this section, we present the mathematical formulation of the hedging problem, along with the computational scheme to obtain the numerical solution.

\subsection{The hedging problem}\label{subse:hedgepro_formulation}
We propose dynamic hedging strategies for managing portfolios of options. Our approach focuses on minimizing a risk measure applied to terminal hedging error while considering variable market conditions and accounting for transaction costs.

The goal is to hedge a short position in a portfolio of contingent claims written on the same underlying asset, $S$, over the hedging period $0,\ldots,T$. The time-$t$ market value of the portfolio is denoted $\mathcal{P}_{t}$. For illustrative purposes, our numerical examples use a European straddle portfolio with maturity $T$. In this case, the value $\mathcal{P}_{T}$ represents the portfolio's terminal payoff, which is given by the mapping $\Psi_T(S_{T})=\max(S_{T}-K,0)+\max(K-S_{T},0)$ with $K$ being the strike price.

The hedging strategy involves managing a self-financing portfolio composed of the risk-free asset, the underlying asset, and a hedging option. Specifically, the hedging option is a European option on the same underlying asset with a longer maturity $T^{*}>T$. The strategy is represented by the predictable process
$\{ \phi_{t} \}_{t=1}^{T}$, with $\phi_{t}=(\phi_{t}^{(r)},\phi_{t}^{(S)},\phi^{(\text{O})}_{t})$, where $\phi_{t}^{(r)}$ is the cash held at time $t-1$ and carried forward to the next period. Moreover, $\phi_{t}^{(S)}$ and $\phi^{(\text{O})}_{t}$ are respectively the number of shares of the underlying asset $S$ and the number of hedging options in the hedging portfolio, both held during the interval $(t-1, t]$. The time-$t$ hedging portfolio value is
\begin{equation*}
V_{t}^{\phi}=\phi_{t}^{(r)}\mbox{e}^{r_{t}\Delta}+\phi_{t}^{(S)} S_{t}\mbox{e}^{q_{t}\Delta} + \phi^{(\text{O})}_{t}\text{O}_{t}(T^{*})
\end{equation*}
where $\text{O}_{t}(T^{*})$ is the time-$t$ hedging option value, $\Delta=\frac{1}{252}$ represents the time increment in years, $r_{t}$ is the time-$t$ annualized continuously compounded risk-free rate and $q_{t}$ is the annualized underlying asset dividend yield, both on the interval $(t-1,t]$. To account for transaction costs the self-financing condition entails that for $t=0,\ldots,T-1$,
\begin{equation}
\label{eq:self-financing}
    \phi_{t+1}^{(r)} +\phi_{t+1}^{(S)}S_{t} + \phi^{(C)}_{t+1}\text{O}_{t}(T^{*})= V_{t}^{\phi}-\kappa_{1} S_{t}\mid \phi_{t+1}^{(S)} - \phi_{t}^{(S)}\mid - \kappa_{2} \text{O}_{t}(T^{*})\mid \phi^{(\text{O})}_{t+1} - \phi^{(\text{O})}_{t}\mid,
\end{equation}
where $\kappa_{1}$ and $\kappa_{2}$ represent the proportional transaction cost rates for the underlying asset and the hedging option, respectively. Transaction costs for options are typically higher than those for the underlying asset. Consequently, we assume $\kappa_{1}<<\kappa_{2}$.

The optimal sequence of actions $\phi= \{ \phi_t\}^T_{t=1}$ corresponds to that which minimizes the application of a risk measure $\rho$ to $\xi_{T}^{\phi}$, the hedging error at maturity for a short position in the option portfolio:
\begin{equation*}
    \xi_{T}^{\phi} = \mathcal{P}_{T} - V_{T}^{\phi}.
\end{equation*}
A positive value in $\xi_{T}^{\phi}$ implies that the hedging strategy does not have enough funds to cover the portfolio value $\mathcal{P}_{T}$. Our goal is to find the hedging strategy $\phi^*$ such that
\begin{equation}\label{Hedging_problem}
\phi^{*}=\mathop{\arg \min}\limits_{\phi} \left\{ \rho  \left( \xi_{T}^{\phi} \right) \right\}.
\end{equation}
Each time-$t$ action $\phi_{t+1}$ is a function of currently available information on the market: $\phi_{t+1}= \tilde{\phi}(X_t)$ for some function $\tilde{\phi}$ of the state variables vector $X_t$. 
Due to Equation \eqref{eq:self-financing}, $\phi_{t+1}^{(r)}$ is fully determined when $\phi_{t+1}^{(S)}$ and $\phi^{(\text{O})}_{t+1}$ are specified, and as such the time-$t$ action to be chosen is $(\phi_{t+1}^{(S)},\phi^{(\text{O})}_{t+1})$.

This paper examines three widely recognized risk measures in the literature:
\begin{itemize}
    \item Mean Square Error (MSE): $\rho\left(\xi_{T}^{\phi}\right)= \mathbb{E}\left[\left(\xi_{T}^{\phi}\right)^{2}\right]$.
    \item Semi Mean-Square Error (SMSE): $\rho\left(\xi_{T}^{\phi}\right)= \mathbb{E}\left[\left(\xi_{T}^{\phi}\right)^{2}\mathbbm{1}_{\{\xi_{T}^{\phi}\geq 0\}}\right]$.
    \item Conditional Value-at-Risk (CVaR$_\alpha$): $\rho\left(\xi_{T}^{\phi}\right)=\mathbb{E}\left[\xi_{T}^{\phi}\middle| \xi_{T}^{\phi}\geq \mbox{VaR}_{\alpha}\left(\xi_{T}^{\phi}\right)\right]$, where $\mbox{VaR}_{\alpha}\left(\xi_{T}^{\phi}\right)$ is the Value-at-Risk defined as $\mbox{VaR}_{\alpha}\left(\xi_{T}^{\phi}\right) = \min_{c}\left\{ c:\mathbb{P}\left(\xi_{T}^{\phi}\leq c\right)\geq \alpha \right\}$, and $\alpha \in (0,1)$.
\end{itemize}

\subsection{Reinforcement learning and deep hedging}\label{subsec:rl_framework}

The problem described in Equation \eqref{Hedging_problem} is addressed by directly estimating the policy function (the investment strategy $\tilde{\phi}$) using a policy gradient method. This approach leverages a parametric representation of the policy function through an Artificial Neural Network (ANN). Specifically, the policy $\tilde{\phi}$, governed by a parameter vector $\theta$, is optimized to minimize the risk measure $\rho$ evaluated at the terminal hedging error. Representing the policy generated by the ANN as $\tilde{\phi}_{\theta}$, the hedging strategy is defined as $\phi_{t+1}=\tilde{\phi}_{\theta}(X_{t})$. Problem \eqref{Hedging_problem} can therefore be approximated as
\begin{equation} 
\label{eq:approxproblem}
    \mathop{\arg \min}\limits_{\theta} \left\{\rho\left(\xi_{T}^{\tilde{\phi}_{\theta}}\right) \right\}. 
\end{equation}
%NO TRADE REGION
Given the inherent continuity of ANNs, the mapping $\phi_{t+1}=\tilde{\phi}_{\theta}(X_{t})$ may lead to frequent small adjustments in the hedging position, potentially increasing long-term transaction costs. To mitigate this effect, we introduce a no-trade region, within which there is no rebalancing. In practice, the no-trade region has a negligible impact on the performance of the ANN, but it improves the results of our benchmark strategies. Further details are provided in \cref{A No trade region}.

As shown in \cite{François2025}, the policy $\tilde{\phi}_{\theta}$ may inadvertently incorporate speculative elements, such as doubling strategies, where agents continuously increase their exposure in an attempt to recover successive losses. Such strategies are undesirable as they deviate from sound risk management principles. To prevent this problem, we introduce a soft tracking error constraint
\begin{equation}
    \label{eq:soft_constraint}
    SC(\theta) = \mathbb{P}\left( \max_{t\in \{0,\dots,T \}} \left\{\xi_{t}^{\tilde{\phi}_{\theta}}\right\} > V_{0} \right)
\end{equation} 
that penalizes the network during training if the time-$t$ tracking error,
\begin{equation}\label{eq:tracking_error}
    \xi_{t}^{\tilde{\phi}_{\theta}} = \mathcal{P}_{t} - V_{t}^{\tilde{\phi}_{\theta}},
\end{equation}
exceeds the initial hedging portfolio value at any time $t$.  
This design does not penalize gains, consistent with the asymmetric nature of rational agents. As a result, instead of solving Problem \eqref{eq:approxproblem}, the objective function employed in our approach is 
\begin{equation}\label{eq:penaltyfunction}
    \mathcal{O}(\theta;\, \lambda) = \rho\left(\xi_{T}^{\tilde{\phi}_{\theta}}\right) + \lambda\, SC(\theta),
\end{equation}
where $\lambda$ is a hyperparameter  that controls the soft constraint weight in the optimization process. It is determined independently using a validation set during the model selection procedure.

We employ a Recurrent Neural Network with a Feedforward Connection (RNN-FNN), integrating Long Short-Term Memory (LSTM) networks with Feedforward Neural Network (FFNN) architectures. This hybrid design has demonstrated superior training performance compared to conventional ANN architectures, as shown in \cite{fecamp2020deep} and \cite{François2025}. The RNN-FNN network is defined as a composition of LSTM cells $\{ C_{l} \}_{l=1}^{L_{1}}$ and FFNN layers $\{ \mathcal{L}_{j} \}_{j=1}^{L_{2}}$ under the following functional representation:
\begin{equation*}
    \tilde{\phi}_{\theta}(X_{t})=(\underbrace{ \mathcal{L}_{J} \circ \mathcal{L}_{L_{2}}\circ \mathcal{L}_{L_{2}-1}\circ ... \circ \mathcal{L}_{1}}_{\mbox{FFNN layers}}\circ \underbrace{C_{L_{1}}\circ C_{L_{1}-1}...\circ C_{1}}_{\mbox{LSTM cells}})(X_{t}).
\end{equation*}
The explicit formulas for this ANN are detailed in \cite{François2025}.

\subsection{Neural network optimization}\label{subse:neural optimization}
 
The RNN-FNN network $\tilde{\phi}_{\theta}(\cdot)$ is optimized with the Mini-batch Stochastic Gradient Descent method (MSGD). This training procedure relies on updating iteratively all the trainable parameters of the optimization problem based on the recursive equations
\begin{equation}\label{updatingrule_1}
    \theta_{j+1} = \theta_{j}-\eta_{j}\frac{\partial}{\partial \theta}\hat{\mathcal{O}}(\theta;\, \lambda),
\end{equation}
%\begin{equation}\label{updatingrule_2}
%    l_{j+1} = l_{j}-\eta_{j}^{(2)}\frac{\partial}{\partial l}\hat{\mathcal{O}}_{\lambda}(\theta,\, l),
%\end{equation}
where $\eta_{j}$ are the learning rates that determine the magnitude of change of parameters per time step. %\footnote{
    %\bc{The no-trade region parameter $\ell$ is updated in a similar manner}.}
These rates are dynamically adjusted using the Adam optimization algorithm.\footnote{
    Adam is an adaptive learning rate method designed to accelerate training in deep neural networks and promote rapid convergence, as detailed in \cite{KingmaB14}.} 
Additionally, $\hat{\mathcal{O}}(\theta;\, \lambda)$ is the Monte-Carlo estimate of the objective function defined by Equation \eqref{eq:penaltyfunction}. Further details can be found in \cref{appen:MSGDtraining}.

%%%%%%%%%%%%%%%%%%%%%%%%%%%%%%%%%%%%%%%%%%%%%%%%%%%%%%%%%%%%%%%%%

\section{Market simulator}\label{se:JMD}

Our approach incorporates a market simulator to emulate the joint dynamics of the S\&P 500 price and of its associated IV surface. Indeed, optimal actions are characterized by the behavior of the underlying asset and the hedging instrument prices. Using a simulator provides the advantage of generating a large diversity of scenarios, enabling RL agents to explore the state space while identifying optimal policies. This alleviates the issue of scarcity in real market data.

We leverage the JIVR model from \cite{Frnacois2023}, which captures the temporal dynamics of S\&P 500 returns alongside the key drivers of the IV surface, while accounting for their interdependencies. The JIVR framework works with interpretable factors and enables the replication of a wide range of realistic IV surface shapes observed in practice.\footnote{Other approaches could be pursued to generate IV surface scenarios, such as generative AI models detailed in \cite{chen2023variational}, \cite{choudhary2024funvol} and \cite{vuletic2024volgan}.} The market simulator has been estimated using a daily dataset of observed implied volatilities—covering a broad range of moneyness and time-to-maturity—alongside S\&P 500 returns from 1996 to 2020; it can therefore reflect a broad array of market conditions. It captures the self-contained properties of the option market, consistently with the "instrumental approach" of option pricing detailed in \cite{rebonato2005volatility}.

\subsection{Daily implied volatility surfaces}
\label{sub:IV_model}

The time-$t$ IV associated to an option with time-to-maturity $\tau_t = \frac{T - t}{252}$ years and (scaled) moneyness \mbox{$M_t = \frac{1}{\sqrt{\tau_t}} \log \frac{S_t e^{(r_t - q_t) \tau_t}}{K}$}
is modeled as
\begin{align}
     & \sigma (M_{t},\tau_{t},\beta_{t}) =  \sum_{i=1}^{5}\beta_{t,i}f_{i}(M_{t},\tau_{t}).
     \label{5factormodel_linear}
\end{align}
The vector $\beta_{t} = (\beta_{t,1}, \beta_{t,2}, \beta_{t,3}, \beta_{t,4}, \beta_{t,5})$ represents the IV factor coefficients at time $t$, while the functions $\{f_{i} \}_{i=1}^{5}$ allow representing the long-term at-the-money (ATM) level, the time-to-maturity slope, the moneyness slope, the smile attenuation, and the smirk, respectively. A detailed description of the functional components $\{f_{i} \}_{i=1}^{5}$ of the IV surface can be found in \cref{subappen:IVsurface}.

\subsection{Joint implied volatility and return}\label{subsec:JIVR}

The JIVR model introduced by \cite{Frnacois2023} builds upon the IV representation \eqref{5factormodel_linear}, offering an explicit formulation for the joint dynamics of the IV surface and the S\&P 500 price. This joint representation is based on an econometric model for (i) the underlying asset returns, and (ii) fluctuations of the IV surface coefficients $\beta_{t}$ along with a mean-reversion component for their volatilities $h_{t}$. The multivariate time series of the JIVR model is provided in \cref{subappen:JIVR_timeseries}.

The JIVR model is used to generate paths of the state variables $(S_{t},\{\beta_{t,i} \}^5_{i=1}, h_{t,R}, \{ h_{t,i}\}^5_{i=1})$, which drive the market dynamics, where $h_{t,R}$ and $\{ h_{t,i}\}^5_{i=1}$ are volatilities for the S\&P 500 and each of the IV factors. 
Estimates of the model parameters and volatility series $\{ \hat{h}_{t,i}\}_{t=1}^{N}$ with $i\in \{1,\ldots,5,R\}$ are taken from \cite{Frnacois2023}.\footnote{
    \cite{Frnacois2023} use a maximum likelihood approach on a multivariate time series made of S\&P 500 returns and surface coefficients estimates $\{ \hat{\beta}_{t}\}_{t=1}^{N}$, with sample dates extending between January 4, 1996 and December 31, 2020.}

%%%%%%%%%%%%%%%%%%%%%%%%%%%%%%%%%%%%%%%%%%%%%%%%%%%%%%%%%%%%%%%%%%%%%%%

\section{Numerical study}\label{se:SSG}

\subsection{Market settings for numerical experiments}\label{subsec:market_parameters}

We consider daily trading periods.
For each simulated path, initial conditions of the JIVR model, $\left(\{ \beta_{0,i} \}^5_{i=1}, h_{0,R}, \{ h_{0,i}\}^5_{i=1}  \right)$, are randomly sampled from the daily estimated values in our data set, covering the period from January 4, 1996, to December 31, 2020. Across all experiments, the annualized continuously compounded risk-free rate and dividend yield are assumed to remain constant, with values fixed at $r = 2.66\%$ and $q = 1.77\%$, respectively.\footnote{The annualized rates of the S\&P 500 dividend yield (1.77\%) and the zero-coupon yield (2.66\%) are calculated as the average over the sample period from January 4, 1996, to December 31, 2020, using OptionMetrics data.}
Without loss of generality, the initial value of the underlying asset is set to $S_0=100$.\footnote{
    In our setting, the value of the portfolio to be hedged is proportional to the underlying asset initial value.} 
The hedged portfolio is an ATM straddle with a maturity of $T=63$ days. At any time $t<T$, the portfolio value \( \mathcal{P}_{t} \) is determined using the IV surface prevailing at that moment. %At maturity $\mathcal{P}_{T}$ represents the final portfolio value, which is the straddle payoff in our example.

The hedging instruments are the risk-free asset, the underlying asset, and an option with a maturity longer than that of the straddle—specifically, an ATM European call option with an initial maturity of $T^{*} = 84$ days. 
%The time-to-maturity of the hedging option naturally decreases over time and is not reset to 84 days at each rebalancing step. 
Positions in all hedging instruments are rebalanced daily.

The hedge follows the self-financing dynamics from Equation \eqref{eq:self-financing}, incorporating proportional transaction costs on both the underlying asset and the hedging option. As reported in \cite{chaudhury2019option}, the average cost for S\&P 500 index call options is 0.95\%. To evaluate its impact, we consider $\kappa_{2} \in \{0.5\%, 1\%, 1.5\%, 2\%\}$. In contrast, transaction costs for the underlying asset are negligible, around 0.047\% according to \cite{BAZZANA2020101240}. We set $\kappa_{1} = 0.05\%$. The initial hedging portfolio value matches the straddle price, \textit{i.e.}, $V_{0}=P_{0}$.

\subsection{Benchmarks}

We benchmark the performance of our framework against several established approaches: (i) the RL method proposed by \cite{François2025}, which incorporates IV-informed decisions using only the underlying asset as a hedging instrument, (ii) delta hedging (D), where only the underlying asset is used for hedging, and (iii) delta-gamma (DG) hedging, which includes the additional hedging option in the portfolio. 

For the second and third benchmarks, the delta and gamma of financial instruments are computed using the \textit{practitioner's} approach, i.e., using the current IV value. 
%inserting the IV for each instrument into the closed-form expressions for Black-Scholes' delta and gamma. 
In the case of delta hedging, the delta is adjusted based on the correction introduced by \cite{Leland}, which accounts for the impact of proportional transaction costs on the underlying asset position. 
%This adjusted delta reverts to the standard Black-Scholes delta when no transaction costs are applied. 
In both benchmarks, the volatility parameter is updated daily according to the prevailing IV surface, which aligns the hedging strategies with dynamic market conditions. The explicit formulas for these two benchmarks are provided in \cref{appen:benchmarks}.

For all three benchmarks, we further enhance the performance by incorporating the no-trade region, as defined in Equation \eqref{eq:association_rule}.\footnote{The optimization process is carried out as detailed in \cref{subse:neural optimization}, following Equation \eqref{updatingrule_1}, using Mini-batch Stochastic Gradient Descent.} Additionally, the no-trade boundary $\ell$ is optimized separately for each risk measure used for benchmarking, with each benchmark exhibiting its own distinct optimal value of $\ell$. 
Further details are provided in \cref{appen:delta_gamma no trade region}.

\subsection{Neural network settings}\label{subse:fine-tuning}

\subsubsection{Neural network architecture}
\label{subsub:network_architecture}
We consider a RNN-FNN architecture with two LSTM cells of width 56, two FFNN-hidden layers of width 56 with ReLU activation function (i.e., $g_{\mathcal{L}_{i}}(X)=\text{max}(0,X)$ for $i=1,2$), and one two-dimensional output FFNN layer with a linear activation function. Numerical experiments detailed in \cref{appen:soft_constraint} from the Supplementary Material suggest the value $\lambda=1$ for the soft constraint  hyperparameter, which is learned from the validation set.

Agents are trained as described in \cref{subse:neural optimization} on a training set of 400,000 independent simulated paths with mini-batch size of 1000 and an initial learning rate of 0.0005.
In addition, we include dropout regularization method with parameter $p=0.5$ as in \cite{François2025}. The training procedure is implemented in Python, using Tensorflow and considering the \cite{glorot2010understanding} random initialization of the initial parameters of the neural network. The performance assessment is obtained from a test set of 100,000 independent paths. 

\subsubsection{State space}

The state space presented in \autoref{tab:state_variables} includes the state variables generated by the JIVR model, along with a new set of state variables associated with the straddle and hedging portfolio. 

\begin{table}[h]
    \centering
    \caption{State variables.}
    \begin{tabular}{>{\arraybackslash}m{2.5cm} >{\arraybackslash}m{8cm}}
        \toprule
        Notation & Description \\
        \midrule
        $S_{t}$ & Underlying asset price \\
        $\{\beta_{t,i} \}^5_{i=1}$ & IV factors described in \cref{sub:IV_model} \\
        $\{ h_{t,i}\}^5_{i=1}$ & IV coefficients' variances \\
        $h_{t,R}$ & Conditional underlying asset return variance \\
        &\\
        $\tau_{t}$ & Time-to-maturity of the straddle \\
        $\mathcal{P}_{t}$ & Straddle value \\
        $\Delta_{t}^{\mathcal{P}}$ & Delta of the straddle \\
        $\Gamma_{t}^{ \mathcal{P}}$ & Gamma of the straddle\\
        &\\
        $\text{O}_{t}$ & Hedging option price\\ 
        $V_{t}^{(\tilde{\phi}_{\theta},\, l)}$ & Hedging portfolio value \\
        $\phi_{t}^{(S)}$ & Underlying asset position \\
        $\phi^{(\text{O})}_{t}$ & Hedging option position \\
        \bottomrule
    \end{tabular}
    \label{tab:state_variables}
    \begin{tablenotes}
    \item \small For all Greeks, as well as the portfolio value and hedging option value, we use the implied volatility $\sigma (M_{t},\tau_{t},\beta_{t})$ from the static surface as the volatility input parameter.
    \end{tablenotes}
\end{table}

In our illustrative example, the RL agent seeks to hedge a straddle contract with the same specifications across different market dynamics. According to the terminology of \cite{peng2024risksensitivecontractunifiedreinforcement}, this problem is a contract-specific reinforcement learning task, where the optimization problem is solved for a given contract with predefined parameters. Variables related to the target portfolio (such as $\mathcal{P}_{t}$, $\Delta_{t}^{P}$, and $\Gamma_{t}^{P}$) are not strictly necessary, as they can theoretically be recovered by the ANN if needed.
However, our numerical experiments demonstrate that in practice their inclusion enhances training performance across all risk measures (details in \cref{appen:state_space}).
Furthermore, incorporating these state variables extends our framework to enable its application in a contract-unified setting, allowing for the optimization of portfolios with any combination of options and contract parameters.

\subsection{Benchmarking of hedging strategies}

\subsubsection{Benchmarking in the absence of transaction costs}\label{subsub:benchmarking_notc}
We begin by evaluating the hedging performance of both benchmark methods and RL agents trained using three different risk measures: MSE, SMSE, and CVaR$_{95\%}$.
This evaluation considers the estimated values of each risk measure alongside the sample average of the hedging error, $$\text{mean}\left(\xi_{T}^{\tilde{\phi}_{\theta}}\right) = \frac{1}{N}\sum_{i=1}^{N} \xi_{T,i}^{\tilde{\phi}_{\theta}},$$ 
where $\xi_{T,i}^{\tilde{\phi}_{\theta}}$ represents the $i$-th terminal hedging error in the test set of size $N$. Additionally, we incorporate the sample standard deviation of the terminal hedging error, $\text{std}\left(\xi_{T}^{\tilde{\phi}_{\theta}}\right)$, as a metric to quantify the variability of hedging errors within the test set. Our analysis is conducted under the assumption of zero transaction costs, i.e., $\kappa_{1}=\kappa_{2}=0$.

\begin{figure}[h]\centering
    \caption{Hedging performance metrics under the assumption of zero transaction costs.}
    \includegraphics[width=16.5cm]{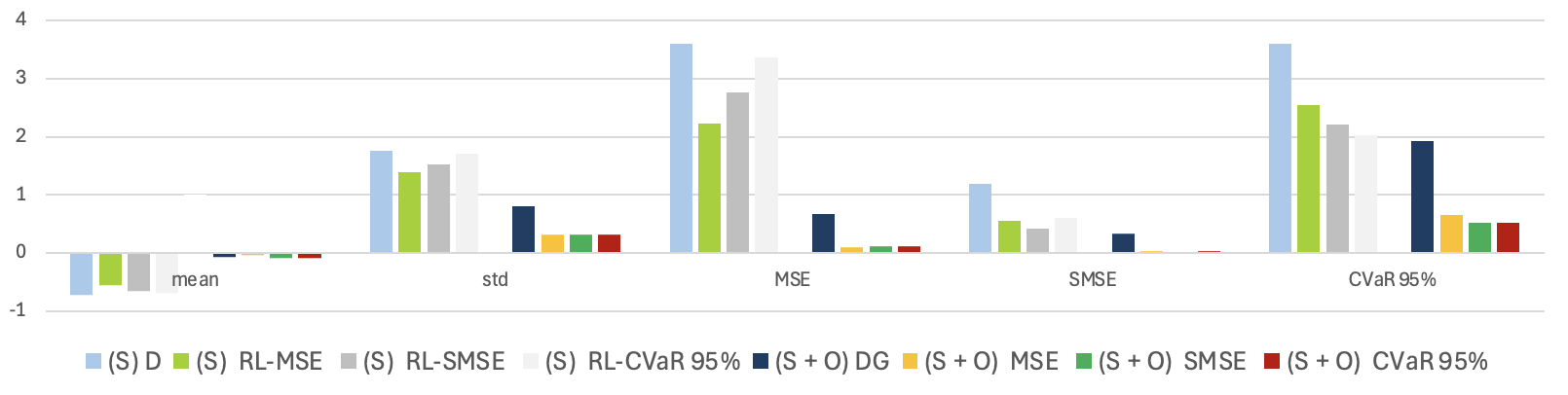}
    \begin{tablenotes}
    \item \small Results are computed using 100,000 out-of-sample paths in the absence of transaction costs ($\kappa_{1}=\kappa_{2}=0$). Agents are trained according to the conditions outlined in \cref{subse:fine-tuning}. The hedged position is an ATM straddle with a maturity of $T = 63$ days and an average value of \$7.55 across all initial conditions. Methods denoted by (S) represent hedging with the risk-free and underlying assets, while those denoted by (S + O) incorporate an ATM call option with an initial maturity of $T^{*} = 84$ days. D stands for delta hedging, DG denotes delta-gamma hedging, and RL refers to reinforcement learning strategies.
    \end{tablenotes}
    \label{fig:aggregated_metrics_notc}
\end{figure}

\autoref{fig:aggregated_metrics_notc} presents the risk measures for the various hedging strategies in two cases. In the first case (the first four columns for each metric), the hedging instruments are limited to the risk-free asset and the underlying asset. In the second scenario (the last four columns), the ATM call option is introduced as an additional hedging instrument. In both cases, RL strategies consistently outperform the benchmarks and achieve the optimal values when the performance assessment metric matches the risk measure used during training.
Our numerical results highlight the benefits of incorporating a second hedging instrument. Specifically, all strategies that include an option as an additional hedging instrument exhibit lower risk in terms of standard deviation, MSE, SMSE, and $\text{CVaR}_{95\%}$, compared to those relying solely on a single hedging instrument. 
%Notably, the delta-gamma hedging strategy reduces standard deviation by at least 42\%, as seen by comparing the lowest standard deviation among single-instrument strategies (1.392) to that of the DG strategy (0.811). It also achieves a 70\% reduction in MSE, a 20\% decrease in SMSE, and a 5\% reduction in CVaR relative to the lowest values of these performance metrics across all strategies in the $S_{t}$ column.
%RL agents utilizing multiple hedging instruments achieve significantly lower risk than the DG strategy, with a standard deviation reduction of at least 60\%—comparing the highest standard deviation among RL strategies with two hedging instruments (0.326) to that of the DG strategy (0.811). This advantage extends to other performance metrics, with reductions of at least 92\% in CVaR$_{95\%}$ and 73\% in SMSE.
Notably, for tail risk captured by CVaR$_{95\%}$, the RL agent—trained on the money account and the underlying asset only—achieves a performance comparable to that of delta-gamma hedging.

\begin{figure}[h]\centering
    \caption{Hedging error distribution in the absence of transaction costs.}
    \includegraphics[width=16.5cm]{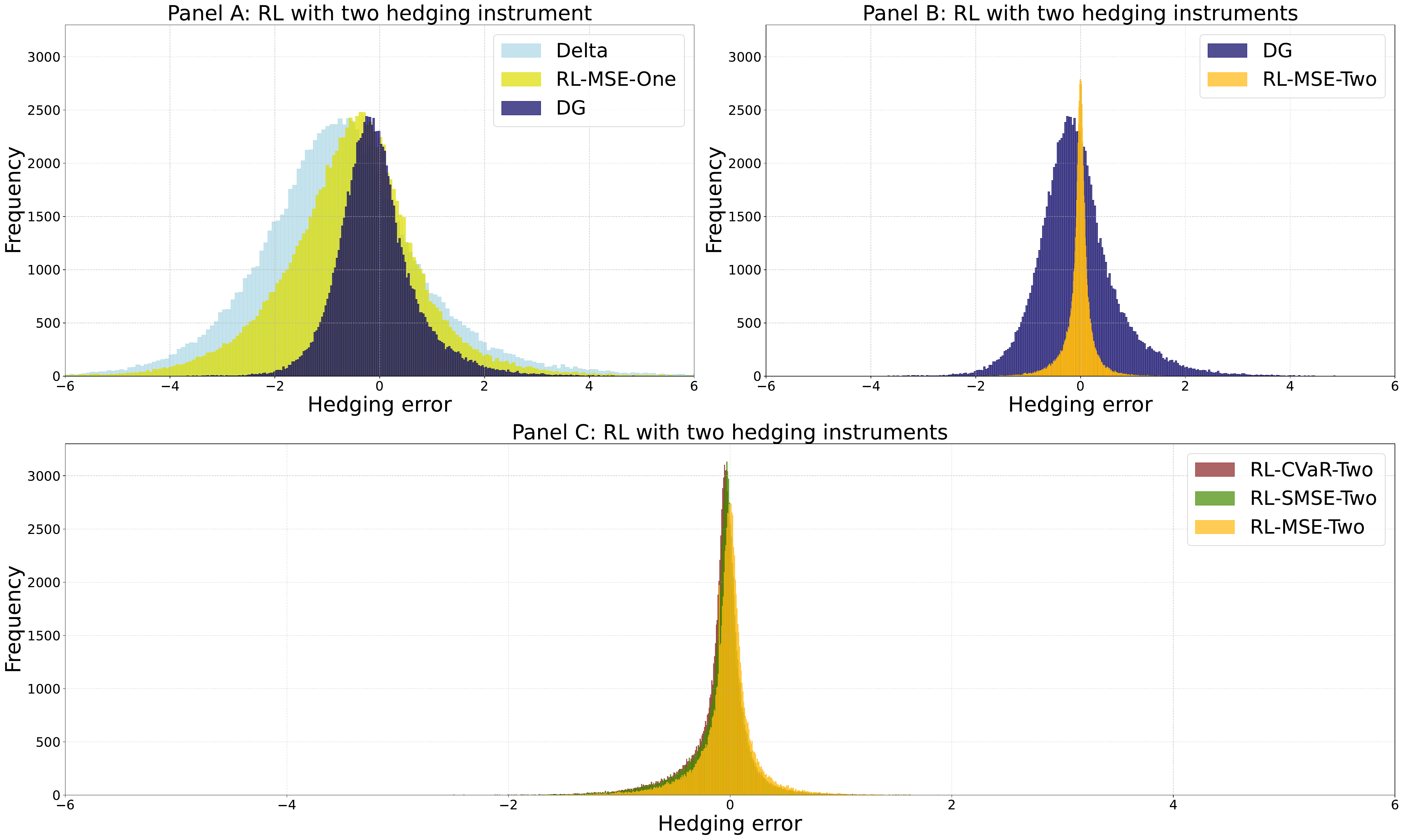}
    \begin{tablenotes}
    \item \small Results are computed using 100,000 out-of-sample paths. The hedged position is an ATM straddle with a maturity of $T = 63$ days and an average value of \$7.55. RL strategies labeled as "One" represent hedging with the risk-free and underlying assets; those labeled as "Two" incorporate an ATM call option with an initial maturity of $T^{*} = 84$ days. 
    \end{tablenotes}
    \label{fig:P&L_distrbution_notc}
    \end{figure}
\autoref{fig:P&L_distrbution_notc} depicts the distribution of hedging errors across various strategies. Panel A contrasts the hedging error distributions of the benchmark and RL agents—both using only the underlying asset—with the traditional DG strategy, showing that incorporating an option significantly reduces risk. Panel B compares the DG strategy to the RL-MSE strategy, both utilizing three hedging instruments, highlighting the RL approach’s superior performance in variance reduction. Finally, Panel C compares the three RL agents, revealing that strategies based on asymmetric risk measures produce distributions with greater skewness.

\subsubsection{Benchmarking in the presence of transaction costs}\label{subsub:benchmarking_tc}
We now measure the impact of transaction costs on the hedging performance.

\begin{figure}[H]\centering
    \caption{Hedging performance in the presence of transaction costs.}
    MSE
    \includegraphics[width=16.5cm]{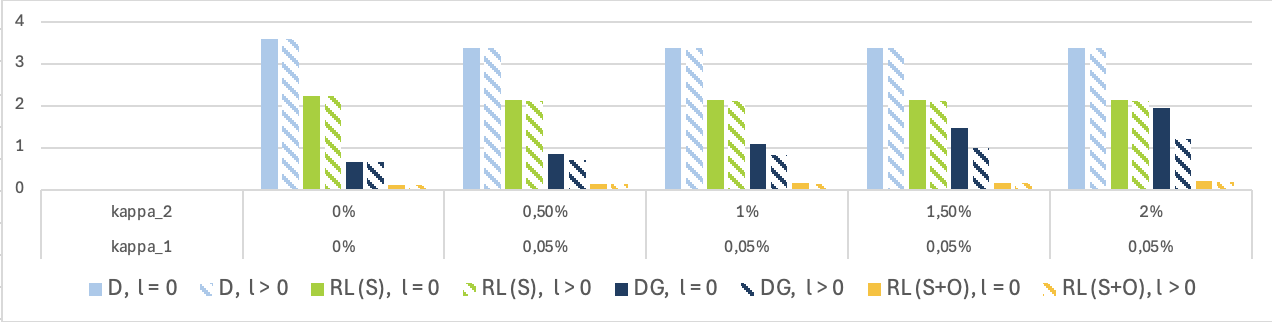} \\
    SMSE
    \includegraphics[width=16.5cm]{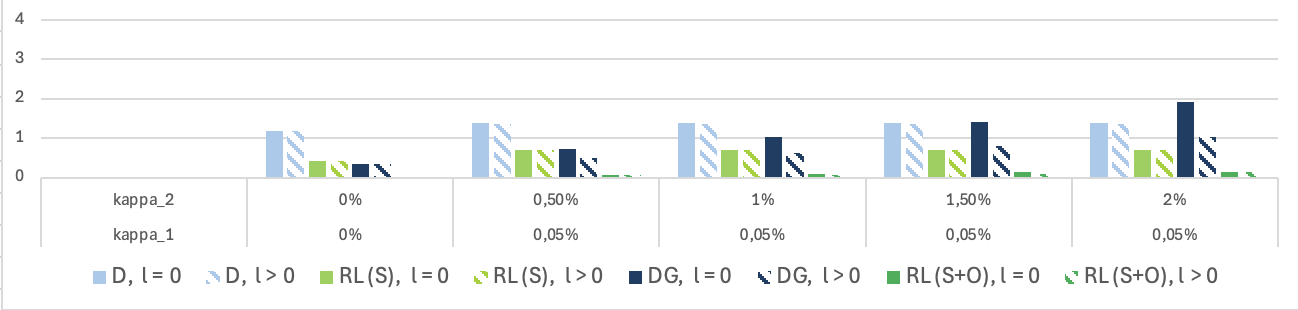}\\
    CVaR
    \includegraphics[width=16.5cm]{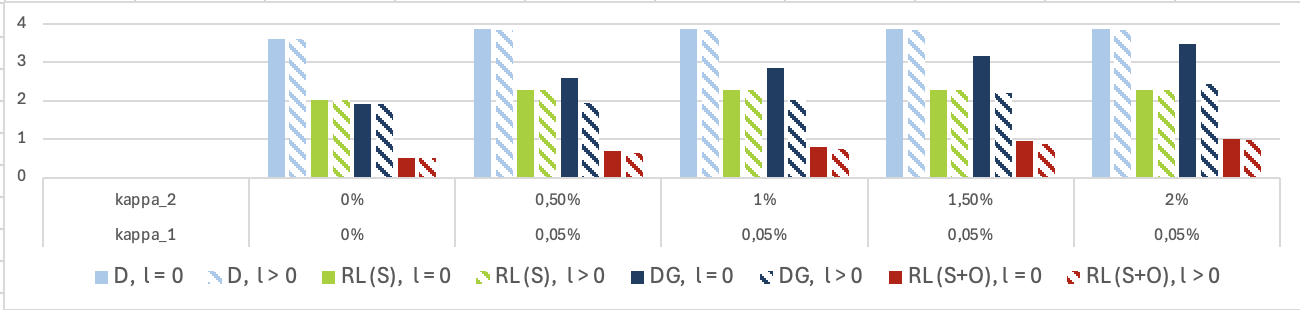}\\
    \begin{tablenotes}
    \item \small Performance metrics are computed using 100,000 out-of-sample paths. The hedged position is an ATM straddle with a maturity of $T = 63$ days and an average value of \$7.55. Methods denoted by (S) represent hedging with the risk-free and underlying assets, while those denoted by (S + O) incorporate an ATM call option with an initial maturity of $T^{*} = 84$ days. D stands for delta hedging, DG denotes delta-gamma hedging, and RL refers to reinforcement learning strategies. Striped bars represent strategies that include the no-trade region, while plain bars correspond to those without it.
    \end{tablenotes}
    \label{fig:hedging_metrics_straddle}
\end{figure}

\autoref{fig:hedging_metrics_straddle} displays the optimal values of risk measures for two distinct hedging configurations: one relying solely on the risk-free asset and the underlying asset (first four columns in each group), and another that includes an ATM call option as an additional hedging instrument (last four columns). The comparison contrasts strategies without a no-trade region (solid bars) against those incorporating a no-trade region (striped bars).
The no-trade region primarily benefits delta-gamma hedging. In the case of delta hedging, transaction costs associated with trading the underlying asset are minimal and have negligible impact on performance. For reinforcement learning (RL) approaches, trading costs are already internalized within the optimization of the neural network policy, rendering the additional constraint of a no-trade region unnecessary. This is consistent with the persistently low threshold values reported in \autoref{fig:optimal_threshold_values} of \cref{A No trade region}.

RL agents consistently outperform benchmarks across all risk measures and choices of hedging instruments. 
Using the MSE as a performance metric, adding an ATM call option as a hedging instrument significantly improves hedging performance. In particular, this holds for the DG strategy, which outperforms RL agents not using options. 
In terms of downside risk management—assessed via SMSE and CVaR metrics—it is noteworthy that, in the presence of transaction costs, the RL algorithm that relies solely on the risk-free asset and the underlying asset either outperforms or provides a performance similar to that of delta-gamma hedging.
RL algorithms that incorporate an option as part of the hedging instruments achieve even stronger performance.

\begin{figure}[H]\centering
    \caption{Hedging error distribution in the presence of transaction costs.}
    \includegraphics[width=16.5cm]{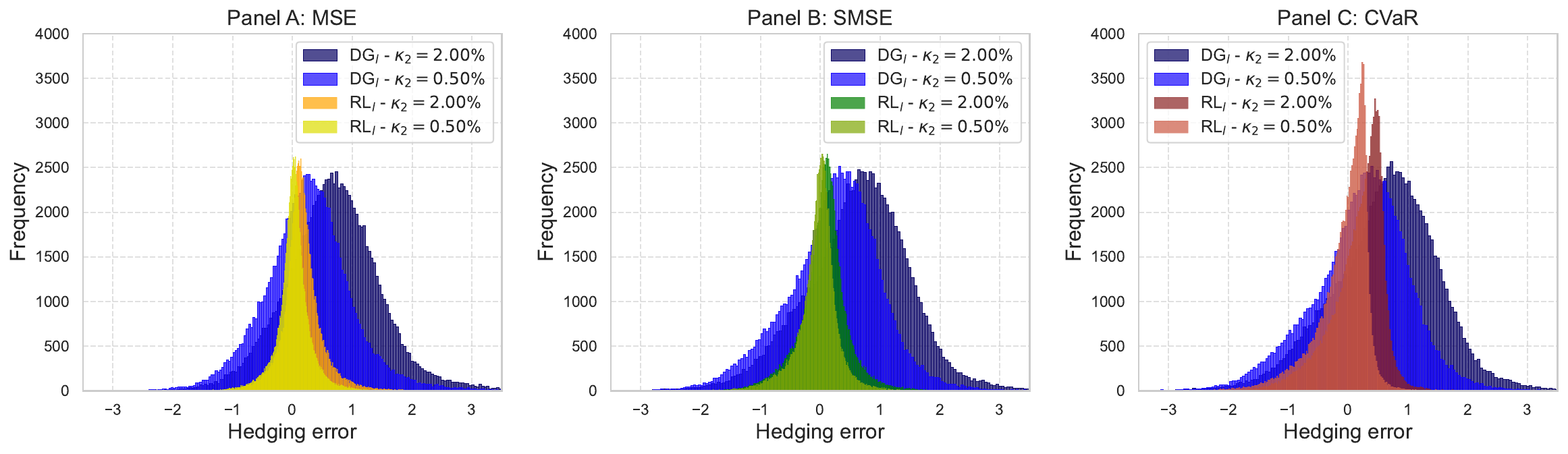}
    \begin{tablenotes}
    \item \small Results are computed using 100,000 out-of-sample paths according to the conditions outlined in \cref{subse:fine-tuning}. The hedged position is an ATM straddle with a maturity of $T = 63$ days and an average value of \$7.55. The hedging instrument is an ATM call option with a maturity of $T^{*} = 84$ days. The transaction cost parameter for the underlying asset is set to $\kappa_{1}=0.05\%$.
\end{tablenotes}
\label{fig:P&L_distrbution}
\end{figure}

To further highlight the advantage of RL over DG, \autoref{fig:P&L_distrbution} presents histograms of hedging error distributions at maturity for both strategies under two different transaction cost scenarios.  RL agents constantly produce narrower distributions across all risk measures, indicating greater resilience to rising transaction costs. This stability is particularly beneficial from a risk management perspective, as it ensures more reliable performance despite increasing costs.

\subsection{Assessing the presence of speculative components in hedging positions}\label{sub:speculative_components}

This section examines whether the RL risk management includes speculative elements, such as strategies that reap the time-varying risk premia embedded in hedging instruments. The risk premium ($\text{RP}$) is defined as the difference between the discounted expected payoff and the option price at time $t$, i.e.,  
\begin{equation}
    \mbox{RP}_{t} = \mbox{exp}(-r(T^{*}-t))\mathbb{E}[\mbox{max}(S_{T^{*}}-K^*,0)\mid \mathcal{F}_{t}] - \mbox{O}_{t}(T^{*})\, ,
\end{equation}
where $K^*$ is the hedging option strike price, the expectation is under the physical measure and $\mathcal{F}_{t}$ denotes the information available at time $t$.\footnote{The usual definition of the risk premium is a return difference. However, when options are DOTM and their value is very low, this definition leads to numerical instability.} The risk premium is estimated using a stochastic-on-stochastic simulation approach, where the present value of the expected payoff is computed through a nested simulation at each time step within the simulated paths.

\begin{figure}[h]\centering
    \caption{Ranked data of risk premium and hedging option positions.}
    \includegraphics[width=16.5cm]{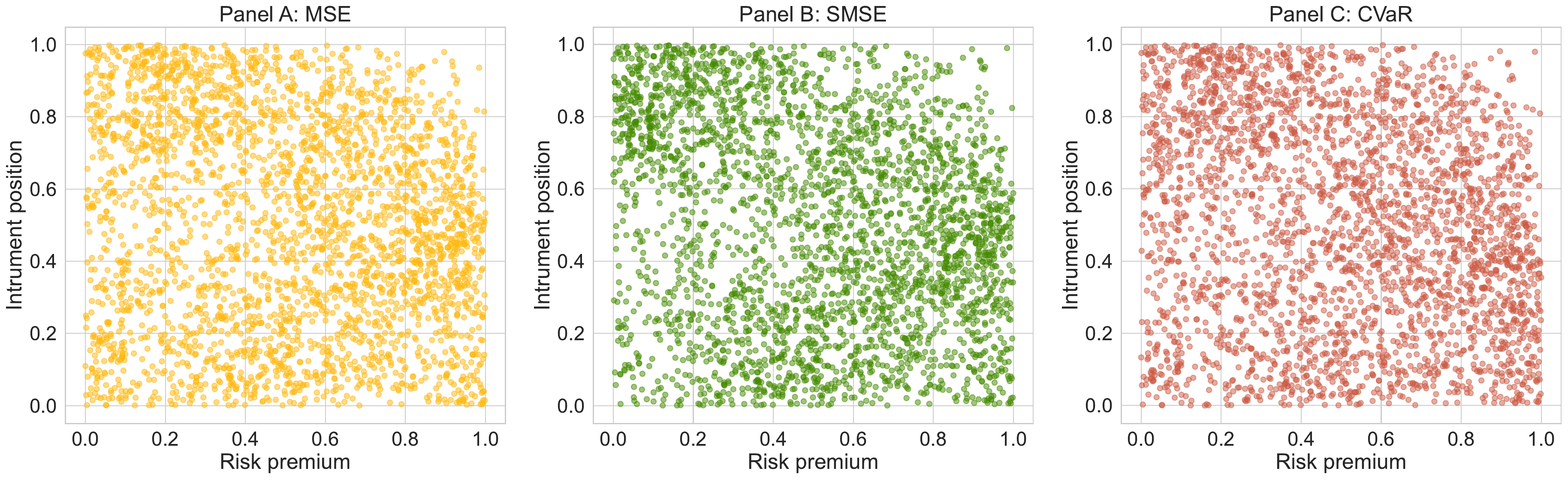}
    \begin{tablenotes}
    \item \small Results are computed using a sample of 20,000 data points from the 100,000 out-of-sample paths. The hedged position is an ATM straddle with a maturity of $T = 63$ days. The hedging instrument is an ATM call option with an initial maturity of $T^{*} = 84$ days. Transaction cost levels are set to 0\%.
    \end{tablenotes}
    \label{fig:risk_premium_dependence}
\end{figure}
We investigate whether a statistical relationship exists between the risk premium $\mbox{RP}_{t}$ and the hedging position $\phi_{t+1}^{(\text{O})}$. \autoref{fig:risk_premium_dependence} presents a scatter plot of ranked data for these variables, using 20,000 samples from the 100,000 out-of-sample paths, which is repeated for the three risk measures. %each risk measure used in the optimization.
The plot reveals no strong dependence patterns, suggesting a weak or insignificant relationship. This finding is further supported by sample correlations ranging from -0.001 to -0.006%across all risk measures
, indicating that RL agents do not systematically seek to capture risk premium benefits.

As a complementary analysis, we examine whether our approach embeds speculative elements, such as statistical arbitrage overlays, that may deviate from sound risk management practices. Our results indicate that RL agents do not engage in such strategies, regardless of the risk measure used in optimization. Further details are provided in \cref{appen:statistical_arbitarge}.

\subsection{Analysis of hedging positions}

\subsubsection{Comparison with benchmarks}

\begin{figure}[h]\centering
    \caption{Pearson correlation between DG and RL agents' hedging option positions.}
    \includegraphics[width=16.5cm]{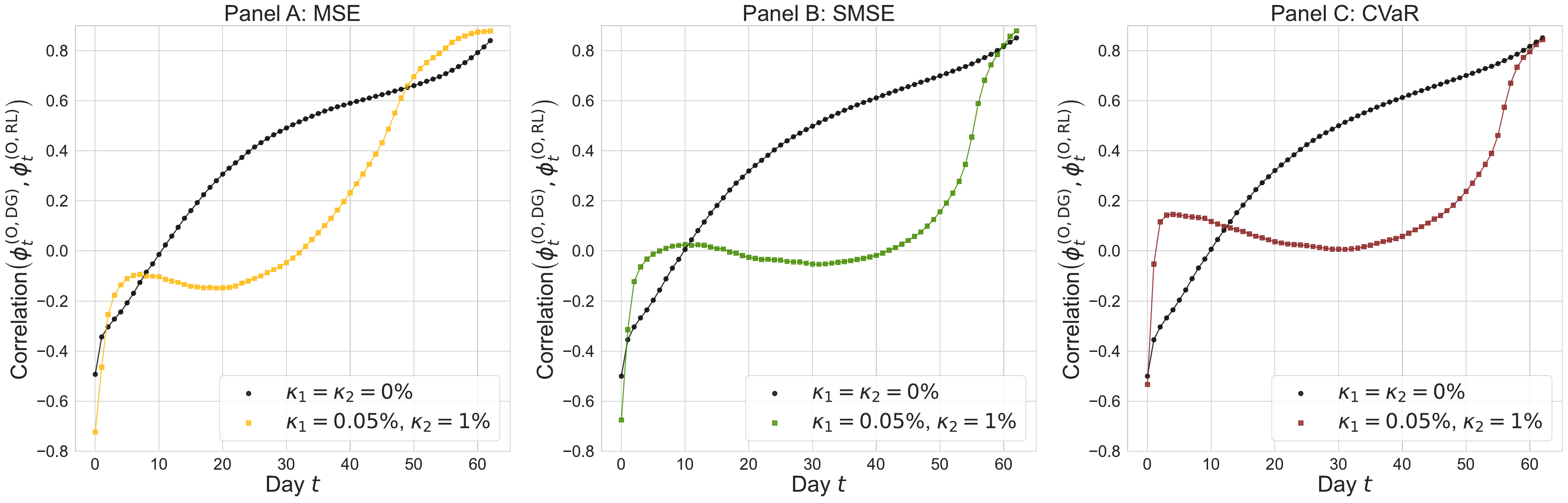}
    \begin{tablenotes}
    \item \small Results are based on a sample of 100,000 out-of-sample paths. Agents are trained under the conditions described in \cref{subse:fine-tuning}. The hedged position is an ATM straddle with a maturity of $T = 63$ day. The hedging instrument is an ATM call option with a maturity of $T^{*} = 84$ days.
    \end{tablenotes}
    \label{fig:pearson_correlation_strategies}
\end{figure}

We analyze the relationship between the hedging option positions produced by the DG strategy and those generated by RL agents. This analysis aims to understand how the RL outperformance documented in \cref{subsub:benchmarking_notc} and \cref{subsub:benchmarking_tc} emerges by studying the positions taken by the hedger. \autoref{fig:pearson_correlation_strategies} presents for various days $t$, the sample correlation between DG and RL hedging option positions, $\phi^{(\text{O},\text{DG})}_t$ and $\phi^{(\text{O},\text{RL})}_t$, under the MSE, SMSE, and CVaR$_{95\%}$ risk measures. 
The correlation is computed for two scenarios: one without transaction costs and another with $\kappa_{1}=0.05\%$ and $\kappa_{2}=1\%$ for illustration.

Our numerical results reveal a consistent pattern across all risk measures, highlighting a significant divergence between RL and DG hedging strategies in terms of correlation, particularly at the start of the hedging horizon.
Indeed, the RL agent benefits from learning experience to anticipate the future movements of state variables over multiple future periods. By contrast, the DG hedging agent is myopic in that he readjusts his hedging positions based on local risk. As time-to-maturity shrinks, both strategies become more similar.
The inclusion of transaction costs leads the RL agent to maintain a distinct approach, with correlation remaining near zero for a significant portion of the hedging horizon.

\begin{figure}[h]\centering
    \caption{Distribution of hedging option positions.}
    \includegraphics[width=16.5cm]{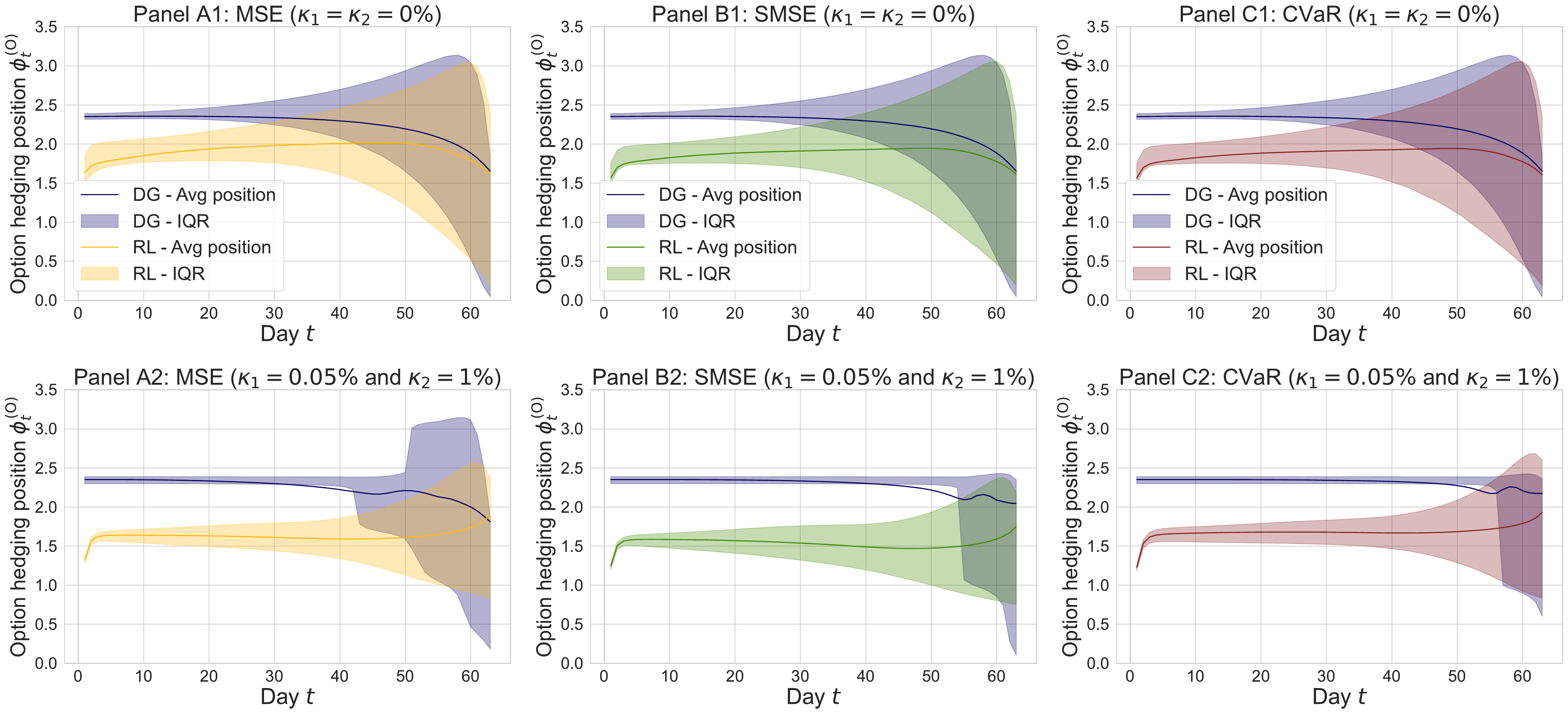}
    \begin{tablenotes}
    \item \small Results are computed over 100,000 out-of-sample paths according to the conditions outlined in \cref{subsub:network_architecture}. The hedged position is an ATM straddle with a maturity of $T = 63$ days. The hedging instrument is an ATM call option with a maturity of $T^{*} = 84$ days. IQR stands for the interquartile range, representing the range between the 25th and 75th percentiles. 
    \end{tablenotes}
    \label{fig:average_position}
\end{figure}

A potential secondary source of divergence between these strategies stems from differences in rebalancing size. 
While the rebalancing frequency influences the timing of adjustments, the magnitude of these adjustments plays a key role in differentiating the hedging behaviors. \autoref{fig:average_position} illustrates the average hedging option position, along with the interquartile range, over time for all risk measures. The analysis is presented for two scenarios: one without transaction costs (first row), and another with transaction costs set to $\kappa_{1}=0.05\%$ and $\kappa_{2}=1\%$ (second row).

Our findings indicate that RL agents tend to hold smaller option positions during the early stages of the hedging period, a trend that is more pronounced with the introduction of transaction costs. This behavior arises from the substantial transaction cost 
associated with the hedging option, suggesting that RL agents favor more frequent rebalancing with smaller initial positions, gradually increasing their hedging positions over time. 
By deferring full engagement with the hedge, the RL agent seeks to balance cost efficiency with effective risk management, avoiding taking positions that might need to be unwound shortly after.
Additionally, lower option positions in early stages allow the agent to initially limit the (short) exposure to the variance risk premium while progressively scaling up the hedging positions. Thus, RL agents achieve twofold cost reductions, where both explicit transaction costs and implicit costs related to short exposure to the variance risk premium are managed.
In contrast, DG strategies adopt larger option positions early in the period to fully neutralize gamma risk. However, this approach leads to prolonged exposure to the volatility premium, making it suboptimal.

\subsubsection{Sensitivity analysis}

We analyze the sensitivity of RL agents' positions to variations in the risk factors defining the IV surface, examining how they leverage information from its shape. Our analysis begins by evaluating RL policy behavior across different initial scenarios for the state variables $(\{\beta_{t,i}\}^5_{i=1}, h_{t,R})$.  

To assess the impact of each state variable, we sort the initial state vectors in the test set according to each variable and observe the corresponding hedging positions in the same order. This method accounts for the interdependence between these state variables and the broader state vector components, as detailed in \autoref{tab:state_variables}, and reveals how changes in a selected variable influence hedging decisions.  
%\gc{We focus on the initial state vector to ensure comparability across market conditions, particularly in terms of the initial underlying asset price and maturity, i.e., at $T=63$ days-to-maturity. GG Not convincing}

\begin{figure}[H]\centering 
    \caption{Impact of state variables on hedging positions.}
    \includegraphics[width=16.5cm]{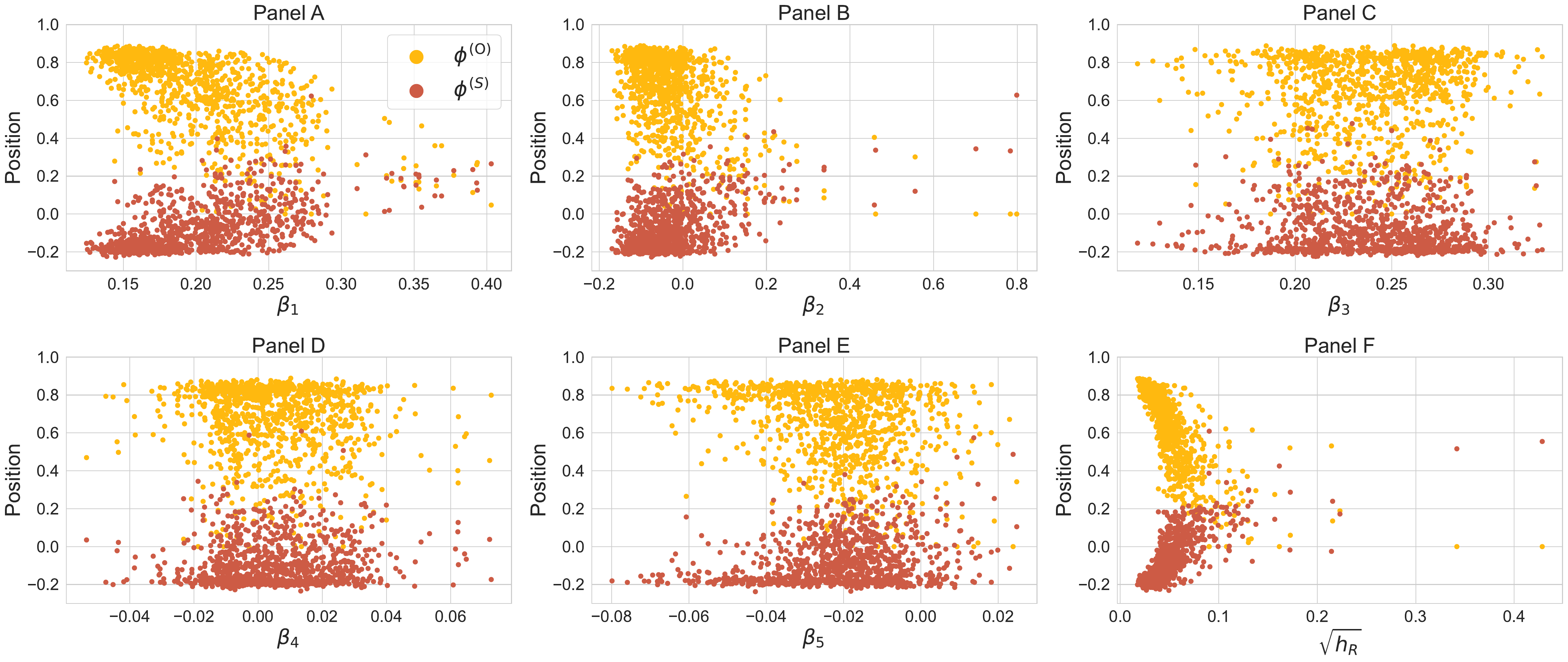}
    \begin{tablenotes}
    \item \small Results are computed using a sample of 20,000 data points from 100,000 out-of-sample paths for an ATM straddle with maturity of $T = 63$ days. The hedging instrument is an ATM call option with an initial maturity of $T^{*} = 84$ days. Transaction cost levels are set to 0\%.
    \end{tablenotes}
    \label{fig:marginal_variation}
\end{figure}

\autoref{fig:marginal_variation} presents the hedging positions of the RL agent trained with the MSE risk measure under a no-transaction-cost scenario. Each panel displays the hedging positions when the initial state vectors are sorted according to each state variable, $(\{\beta_{t,i}\}^5_{i=1}, h_{t,R})$.

These empirical results suggest that the position in the hedging option exhibits a decreasing trend with respect to the conditional variance of the underlying asset returns, the long-term ATM level $\beta_1$ and the time-to-maturity slope $\beta_2$ of the IV surface.\footnote{By contrast, there is no clear pattern related with the other factors as shown in panels C, D and E.} As noted in \cite{François2025}, RL agents utilize both the historical variance process and market expectations of future volatility to adjust their positions. For instance, smaller positions on the hedging option when $\beta_1$, $\beta_2$ or $\sqrt{h_{R}}$ are higher can be explained by the higher cost of hedging in such circumstances. Indeed, both option prices and associated proportional transaction costs are higher.

\subsection{Tracking error analysis}

The differences between positions of RL and DG agents allow RL agents to achieve higher performance with respect to terminal hedging error. This section investigates whether RL agents
also retains good tracking performance before maturity.

We analyze the time-$t$ tracking error $\xi_{t}^{\tilde{\phi}_{\theta}}$ defined in Equation \eqref{eq:tracking_error} across all test set paths %for different strategies 
throughout the hedging period. 
This comparison is conducted by evaluating three key metrics on each rebalancing day $t$: the average tracking error (ATE), root-mean squared tracking error (RMSTE), and semi root-mean squared tracking error (SRMSTE),
given respectively by  
\begin{equation*}
    \text{ATE}=\frac{1}{N} \sum_{i=1}^{N} \xi_{t,i}^{(\tilde{\phi}_{\theta},\, l)}, \,
    \text{RMSTE}=\sqrt{\frac{1}{N} \sum_{i=1}^{N} \left(\xi_{t,i}^{(\tilde{\phi}_{\theta})}\right)^2}, \,
    \text{SRMSTE}=\sqrt{\frac{1}{N} \sum_{i=1}^{N} \left(\xi_{t,i}^{(\tilde{\phi}_{\theta})} \mathbbm{1}_{\left\{ \xi_{t,i}^{(\tilde{\phi}_{\theta})} > 0 \right\}}\right)^2},
\end{equation*}
where $\xi_{t,i}^{\tilde{\phi}_{\theta}}$ represents the time-$t$ tracking error of the $i$-th path in the test set.

\begin{figure}[h]\centering 
    \caption{Evolution of tracking error metrics across rebalancing days.}
    \includegraphics[width=16.5cm]{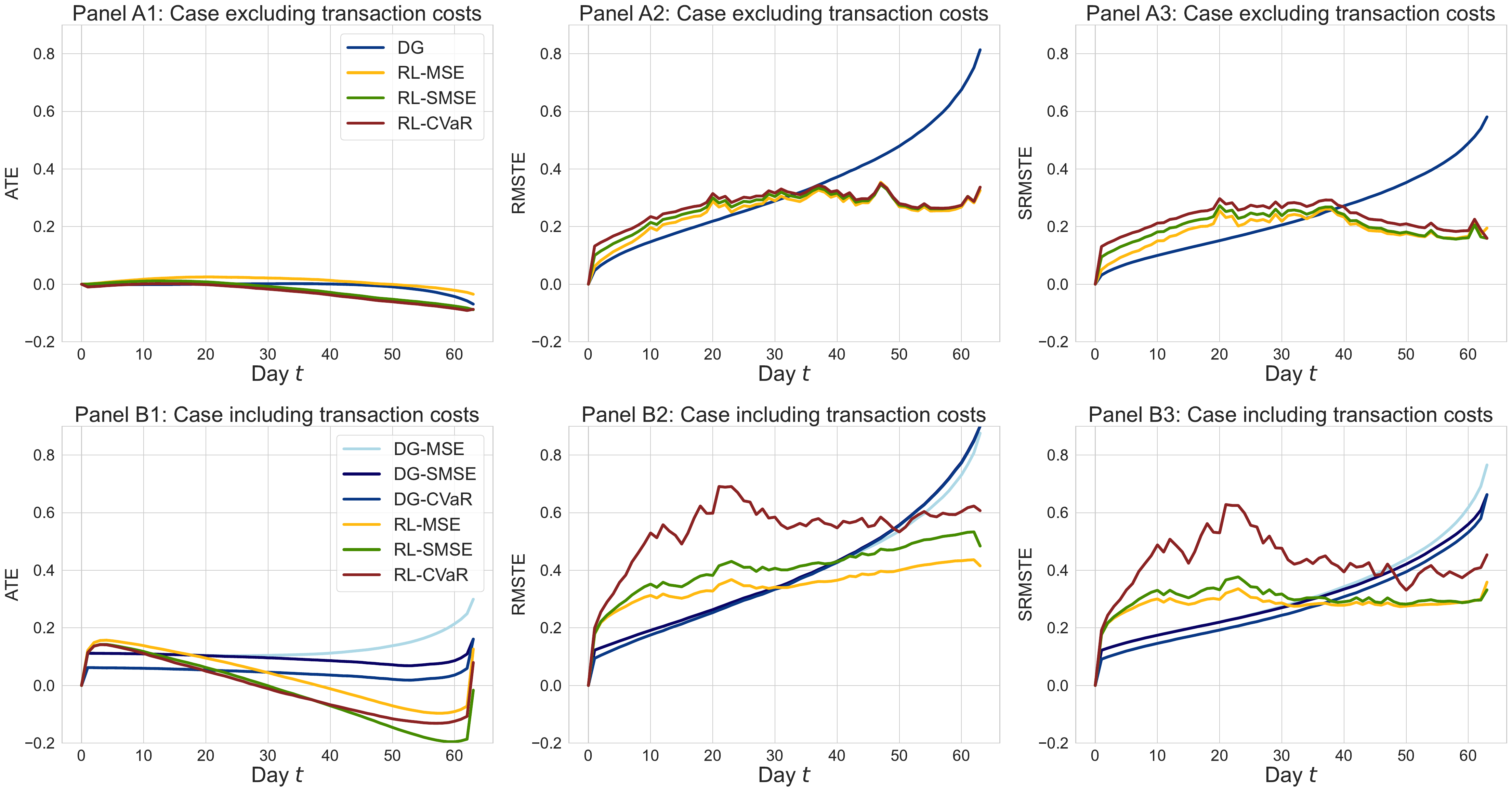}
    \begin{tablenotes}
    \item \small Results are computed over 100,000 out-of-sample paths under the conditions outlined in \cref{subsub:network_architecture}. The hedged position is an ATM straddle with a maturity of $T = 63$ days and an average value of \$7.55. The hedging instrument is an ATM call option with a maturity of $T^{*} = 84$ days.
    \end{tablenotes}
    \label{fig:tracking_error}
\end{figure}

\autoref{fig:tracking_error} presents the evolution of these metrics over the hedging period under two scenarios: without transaction costs (Panel A) and with transaction costs (Panel B). Panel B accounts for multiple DG strategies, each corresponding to a different optimal no-trade threshold $\ell$. 
The results indicate that, regardless of transaction costs, both the standard and asymmetric tracking error metrics (columns 2 and 3 of \autoref{fig:tracking_error}) exhibit a monotonic upward trend for DG strategies. In contrast, RL strategies lead to curves that flatten out or even decrease through time, demonstrating their ability to correct for past errors. Conversely, DG strategies are purely forward-looking, leading to the accumulation of unaddressed errors over time.

Furthermore, columns 2 and 3 show that RL agents maintain strong option-tracking performance in the absence of transaction costs, despite adopting strategies that differ from those derived using the DG approach. However, once transaction costs are introduced (panels B2 and B3 of \autoref{fig:tracking_error}), the RL agent trained under the CVaR risk measure exhibits larger tracking error. This is primarily driven by the nature of the objective function, which focuses on minimizing the tail of losses only at the end of the hedging period. As a result, early deviations between the hedging and target portfolios 
do not necessarily lead to a loss in the tail of the distribution, and therefore do not require immediate correction, as positions can be rebalanced closer to maturity while keeping the CVaR at low levels. Larger tracking errors in early stages are expected because the RL optimization leads to smaller hedging positions, see \autoref{fig:average_position}.

In terms of the sample average tracking error (column 1), DG strategies exhibit values close to zero across all rebalancing days in absence of transaction costs. The RL agent trained under the MSE risk metric follows closely, which aligns with the symmetric nature of this risk measure, as it penalizes both losses and gains equally.  
In contrast, RL strategies optimized using SMSE and CVaR deviate further from zero, particularly displaying a negative average hedging error. This behavior reflects the asymmetric nature of these risk metrics, which do not penalize gains. These differences become even more pronounced when transaction costs are introduced, further emphasizing the distinct risk preferences embedded in each optimization approach.

\section{Out-of-sample backtesting} \label{subsub:benchmarking_realpaths}

We assess the performance of our framework under actual market conditions, using historical option prices sourced from OptionMetrics observed between December 31, 2020, and October 31, 2023. We evaluate the hedging performance across 4,134 near-the-money 63-day European straddles, where the option strike lies within $\pm10\%$ of the underlying asset's initial price. Each straddle is hedged using a combination of a call option with a longer maturity (between 78 and 84 days of maturity depending on availability), the underlying asset and the cash account.\footnote{On each day of the out-of-sample dataset, the IV parameters $\beta_t$ are estimated.  We then re-estimate their joint dynamics with the S\&P 500 returns to recreate the state space. To ensure consistency with the simulation environment used during training, all underlying price paths are rescaled to start at a normalized value of 100.}

\begin{figure}[H]\centering
    \caption{Out-of-sample backtest performance metrics on hedging errors, with and without transaction costs.}
    \includegraphics[width=16.5cm]{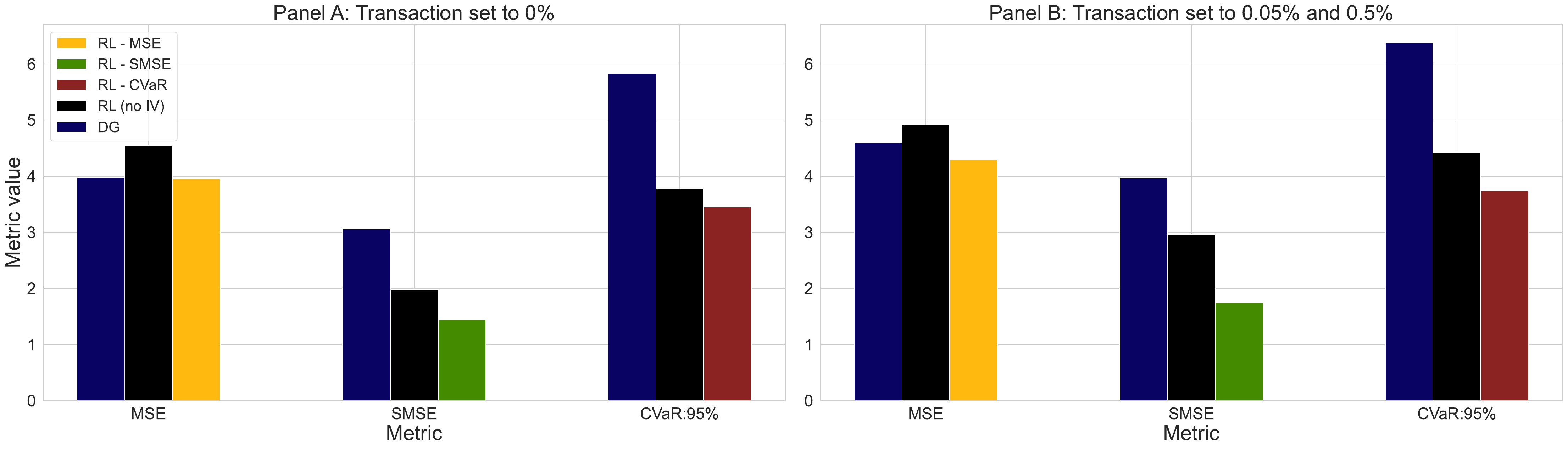}
    \begin{tablenotes}
    \item 
    \small
    The backtest is conducted on 4,134 around-the-money straddle intruments, using actual market prices observed between December 31, 2020 and October 31, 2023. %The  average value of the hedged position is \$9.12.
    \end{tablenotes}
    \label{fig:backtest_metrics_comparison}
\end{figure}

We compare the performance of the practitioners’ delta-gamma hedging with that of the RL algorithms with and without IV surface information.\footnote{The RL benchmark without IV surface information is analogous to the proposed RL method, except that predictors $\beta_{t,1},\ldots,\beta_{t,5},h_{t,1},\ldots,h_{t,5}$ are dropped from the state space.}
\autoref{fig:backtest_metrics_comparison} shows that, in terms of MSE, the RL algorithm without IV surface information performs worse than the two other approaches. Interestingly, in the absence of transaction costs, the practitioners’ delta-gamma hedging and the RL algorithm with IV information exhibit very similar MSE. The RL-algorithm with the complete information slightly outperforms the other approaches in presence of transaction costs, again in terms of MSE. The main conclusion is that RL approaches do not necessarily dominate traditional methods; their performance critically depends on the information provided to the algorithm.
However, in terms of tail risk, the RL algorithms clearly outperform the practitioners’ delta-gamma approach. Moreover, receiving information about the IV surface clearly improves the performance the RL algorithm.

\begin{figure}[h]\centering
    \caption{Distribution of hedging errors for near-at-the-money straddles.}
    \includegraphics[width=16.5cm]{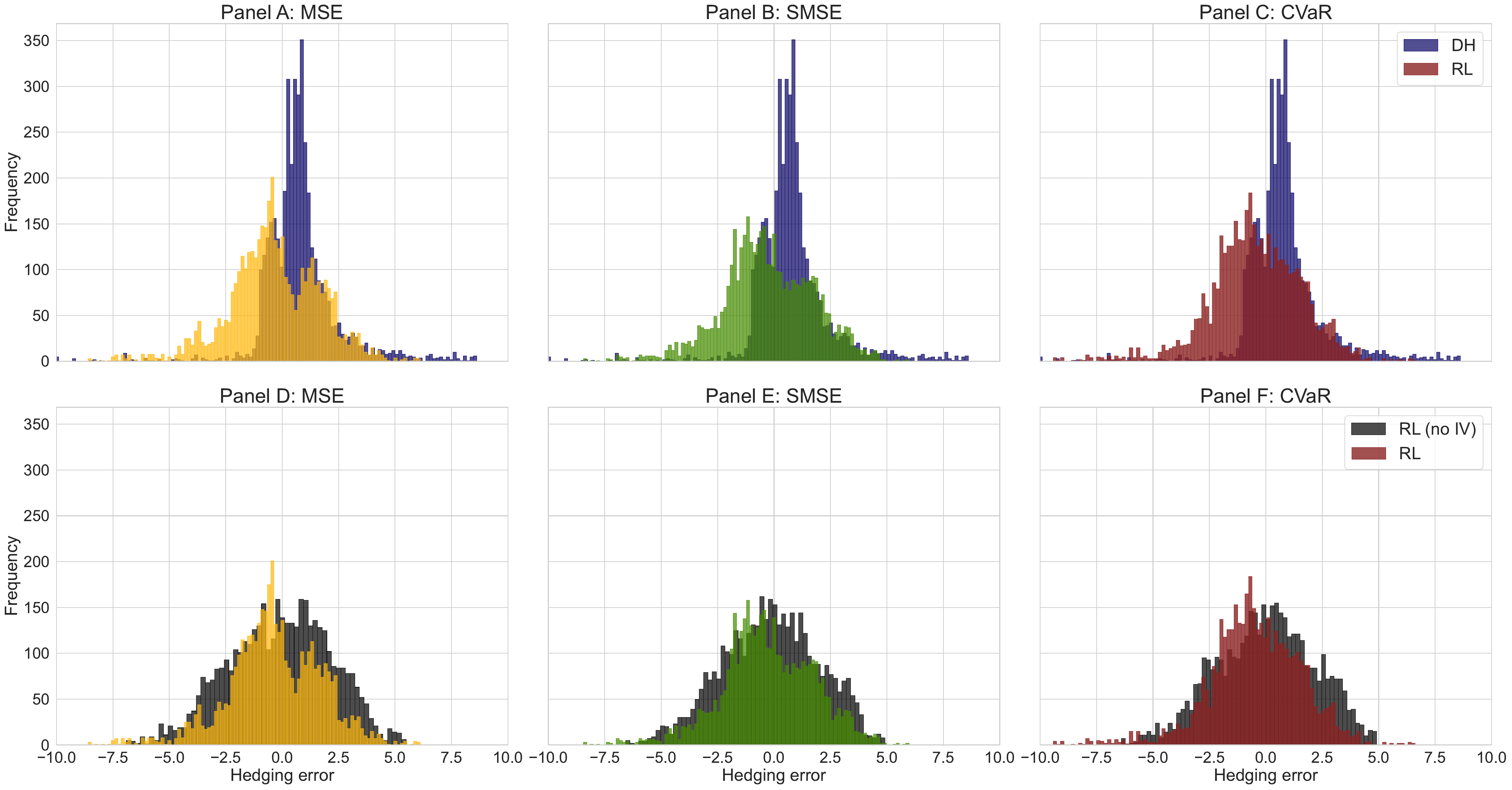}
    \begin{tablenotes}
    \item \small The backtest is conducted on 4,134 around-the-money straddle intruments, using actual market prices observed between December 31, 2020 and October 31, 2023. No transaction costs are applied.
    \end{tablenotes}
    \label{fig:backtest_distribution_f}
\end{figure}

Further evidence of the RL approach's superior performance is provided by the distribution of hedging errors (without transaction costs) shown in \autoref{fig:backtest_distribution_f}. The first row compares the RL algorithm with the full information to the practitioners’ delta-gamma strategy. It is clear that the distribution of the practitioners’ delta-gamma strategy is shifted to the right and exhibits a heavier right tail.
In the second row, we observe the benefits of incorporating IV information into the RL strategies. The distribution of hedging errors is more concentrated around zero when such information is included.

\begin{figure}[h]\centering
    \caption{Time series of hedging strategies P\&L and market conditions.}
    \includegraphics[width=16.5cm]{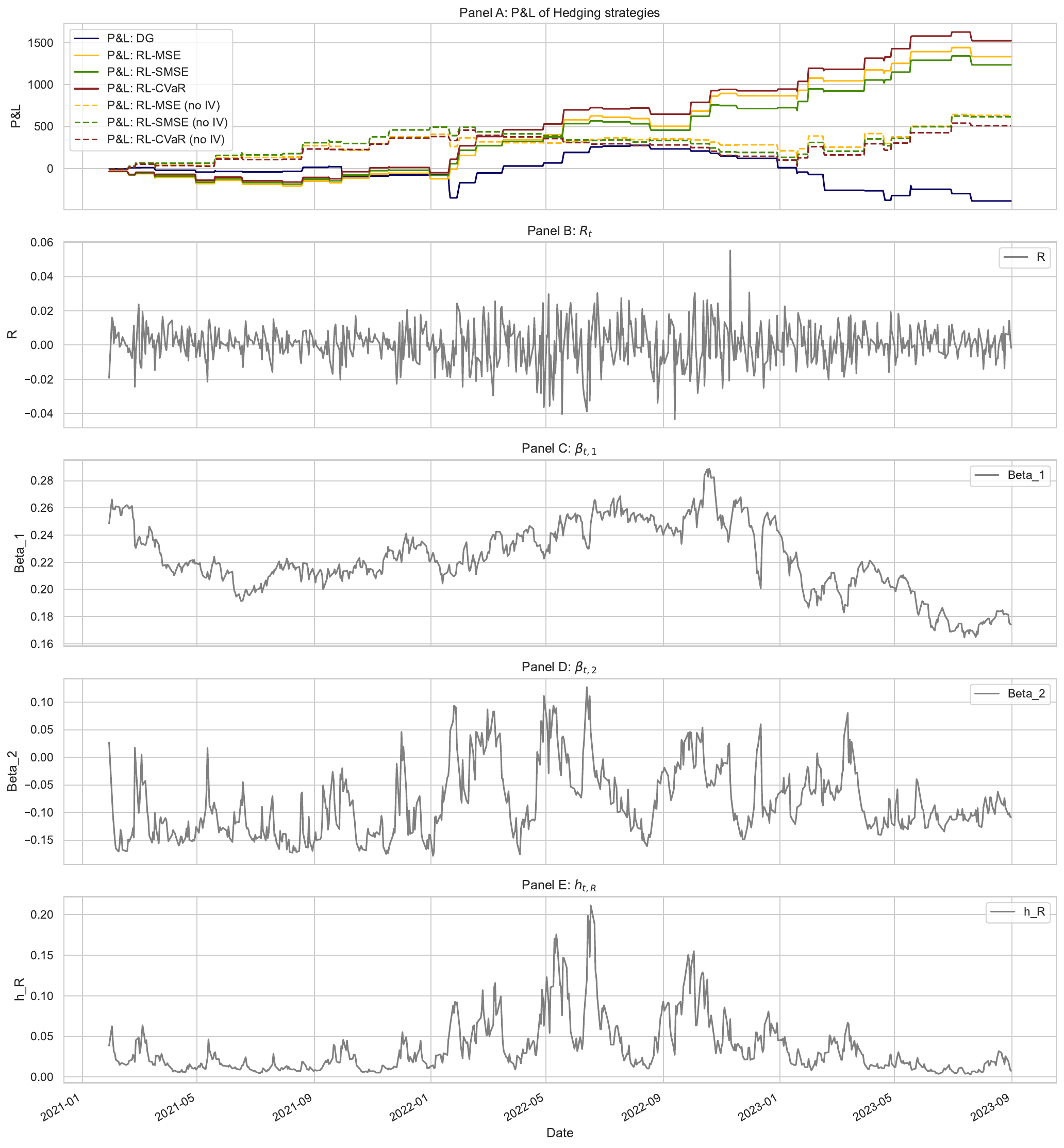}
    \begin{tablenotes}
    \item 
    \small The backtest is conducted on 4,134 around-the-money straddle intruments, using actual market prices observed between December 31, 2020 and October 31, 2023. 
    %The average value of the hedged position is \$5.39. 
    No transaction costs are applied.
    \end{tablenotes}
    \label{fig:backtest_timeseries}
\end{figure}
Although risk management is primarily associated with measures of dispersion and downside risk, it is interesting to note that only the practitioners’ delta-gamma strategy exhibits a negative cumulative P\&L (see \autoref{fig:backtest_timeseries}). Moreover, having the full information about the IV surface helps the RL algorithm achieve the best cumulative P\&L. We observe that the cumulative P\&L of the RL algorithm with the full information increases significantly in the second half of the sample. Examining market conditions (Panel B to Panel E), we see that this period is associated with IV slopes that are strongly positive (see Panel D). In the middle of the sample, there is a period of high volatility (see Panel E), but this information is captured by both the practitioners’ delta-gamma hedging and the RL algorithms.

These findings demonstrate that RL agents achieve consistent and competitive performance when applied to unseen historical market conditions, despite being trained on simulated data. Their ability to adapt to diverse environments and maintain superior risk control highlights the practical value of this approach in hedging tasks.

%%%%%%%%%%%%%%%%%%%%%%%%%%%%%%%%%%%%%%%%%%%%%%%%%%%%%%%%%%%%%%%%%%%%%%%%%
%%%%%%%%%%%%%%%%%%%%%%%%%%%%%%%%%%%%%%%%%%%%%%%%%%%%%%%%%%%%%%%%%%%%%%%%%
%%%%%%%%%%%%%%%%%%%%%%%%%%%%%%%%%%%%%%%%%%%%%%%%%%%%%%%%%%%%%%%%%%%%%%%%%
%%%%%%%%%%%%%%%%%%%%%%%%%%%%%%%%%%%%%%%%%%%%%%%%%%%%%%%%%%%%%%%%%%%%%%%%%

\section{Conclusion}\label{se:conclusion}

This study develops
a deep hedging framework to manage the risk associated with S\&P 500 options with a hedging portfolio including both options and underlying asset shares. In our work the information related to implied volatility surfaces is included within the set of state variables. 
The key differentiating aspect of our work is that with this information in hand, the adjustments in hedging positions not only integrate forward-looking expectations of market dynamics, but also capture the current price levels for options (and the associated variance risk premium) within rebalancing decisions. The IV surface, conveniently represented by a parametric form, proves to be instrumental in refining the hedging policy. A soft constraint is included in the optimization scheme to mitigate speculative behavior, ensuring that hedging strategies focus on effective risk management.

%Our hedging framework is also enhanced with a couple of trading features: (i) state-dependent no-trade regions to optimize rebalancing frequency in the presence of transaction costs, (ii) a soft constraint to mitigate speculative behavior, ensuring that hedging strategies focus on effective risk management.

Our approach consistently outperforms traditional benchmarks both with and without transaction costs. It also highlights the substantial hedging benefits of incorporating additional instruments, such as options. 
%Furthermore, the inclusion of no-trade regions improves performance for both reinforcement learning and delta-gamma strategies: The former reduces unnecessary rebalancing, while the latter behaves like semi-static hedging approaches.
Our study further documents the reasons driving the hedging outperformance of the reinforcement learning agent. In contrast to the myopic delta-gamma hedging, deep hedging begins with smaller option positions. This leads to less transaction costs and, more importantly, provides more flexibility for appropriately rebalancing the hedging portfolio when uncertainty about the final moneyness of the position to hedge is gradually resolved.
Smaller early-stage positions in the hedging option also reduce exposure to the variance risk premium, leading to lower losses.
We show that reinforcement learning agents effectively incorporate both historical variance and market expectations of future volatility into their hedging decisions. The observed decline in hedging option positions in response to higher conditional variance, long-term ATM implied volatility level and time-to-maturity slope underscores the agents' ability to dynamically mitigate risk, acting as a protective mechanism against volatility fluctuations.

Out-of-sample backtests using historical data and various levels of transaction costs show that the reinforcement learning hedging performance is robust to diverse market conditions and superior to that of benchmarks in terms of downside risk management, on top of providing superior profitability. Such tests highlight the importance of information embedded in implied volatility surfaces. This confirms that deep hedging with options using the implied volatility surface is a sound and practically applicable hedging approach.

%%%%%%%%%%%%%%%%%%%%%%%%%%%%%%%%%%%%%%%%%%%%%%%%%%%%%%%%%%%%%%%%%%%%%%%%%
%%%%%%%%%%%%%%%%%%%%%%%%%%%%%%%%%%%%%%%%%%%%%%%%%%%%%%%%%%%%%%%%%%%%%%%%%
%%%%%%%%%%%%%%%%%%%%%%%%%%%%%%%%%%%%%%%%%%%%%%%%%%%%%%%%%%%%%%%%%%%%%%%%%
%%%%%%%%%%%%%%%%%%%%%%%%%%%%%%%%%%%%%%%%%%%%%%%%%%%%%%%%%%%%%%%%%%%%%%%%%

\bibliographystyle{apalike}
%\bibliographystyle{abbrvnat}
%\bibliographystyle{unsrtnat}
% plain
\bibliography{references}  

%\newpage

%--------
\appendix
%--------
\section*{Appendices}

\section{No trade region}
\label{A No trade region}

At time $t$, the no-trade region\footnote{
    No-trade regions, which mitigate the impact of transaction costs, have been extensively studied in the portfolio optimization literature. \cite{george_capital_cost} first introduced the idea that proportional transaction costs give rise to such regions—a concept further developed by \cite{davis_norman_1990} and \cite{balduzzi1999transaction}, who emphasized portfolio allocation over rebalancing costs. In the hedging context, optimal rebalancing based on delta variations has been explored by \cite{henrotte1993transaction}, \cite{toft1996mean}, and \cite{martellini2002competing}. \cite{hodges1989optimal} and \cite{clewlow1997optimal} examine hedging within a utility-maximization framework. The optimal hedging strategy consists of no-trade bands around delta, whose width depends on the hedger's risk aversion.
    }  is determined by the distance between the current portfolio position, $\phi_t$, and the next position proposed by the ANN, $\tilde{\phi}_{\theta}(X_t)$. Specifically, rebalancing occurs only if the cumulative deviation in positions across hedging instruments exceeds a threshold $l$:
\begin{align}\label{eq:association_rule}
    (\phi^{(S)}_{t+1}, \phi^{(O)}_{t+1}) &= 
    \begin{cases} 
        (\phi^{(S)}_{t}, \phi^{(O)}_{t}), & \text{if } \lvert \phi_{t}^{(S)} - \tilde{\phi}_{\theta}^{(S)}(X_{t}) \rvert + \lvert \phi^{(\text{O})}_{t} - \tilde{\phi}_{\theta}^{(\text{O})}(X_{t}) \rvert \leq \ell, \\[8pt]
        \left(\tilde{\phi}^{(S)}_{\theta}(X_{t}),\, \tilde{\phi}^{(O)}_{\theta}(X_{t}) \right), & \text{otherwise}.
    \end{cases}
\end{align}
The bank account position is determined by the self-financing constraint \eqref{eq:self-financing}. This formulation expresses the no-trade region in terms of the number of shares of option contracts, providing a measure of the distance at which rebalancing becomes cost-effective, capturing the trade-off between transaction costs and maintaining proximity to the desired portfolio adjustments.
Indeed, when rebalancing actions proposed by the neural network are minor, they are not implemented because (i) this only leads to a small misalignment with the ideal hedging positions and (ii) this allows avoiding transaction costs.\footnote{We tried other specifications for the no-trade region (for instance explicitly capturing transaction cost amounts), with results being qualitatively similar.}
The rebalancing threshold $\ell$ is treated as a learnable parameter included in the ANN parameters $\theta$, allowing the model to jointly optimize the size of rebalancing actions and decisions of whether or not to rebalance.

%\subsection{Numerical results}
%\begin{equation}\label{updatingrule_2}
%    l_{j+1} = l_{j}-\eta_{j}^{(2)}\frac{\partial}{\partial l}\hat{\mathcal{O}}_{\lambda}(\theta,\, l),
%\end{equation}
This analysis incorporates the no-trade region, defined by Equation \eqref{eq:association_rule}, to optimize rebalancing frequency while accounting for transaction costs. 
For benchmarks, the rebalancing threshold $\ell$ is estimated using the approach described in \cref{appen:delta_gamma no trade region}. In contrast, RL strategies estimate this parameter jointly with other ANN parameters during training. 

\begin{figure}[H]\centering
    \caption{Optimal rebalancing threshold $\ell$ values for DG and RL strategies.}
    \includegraphics[width=16cm]{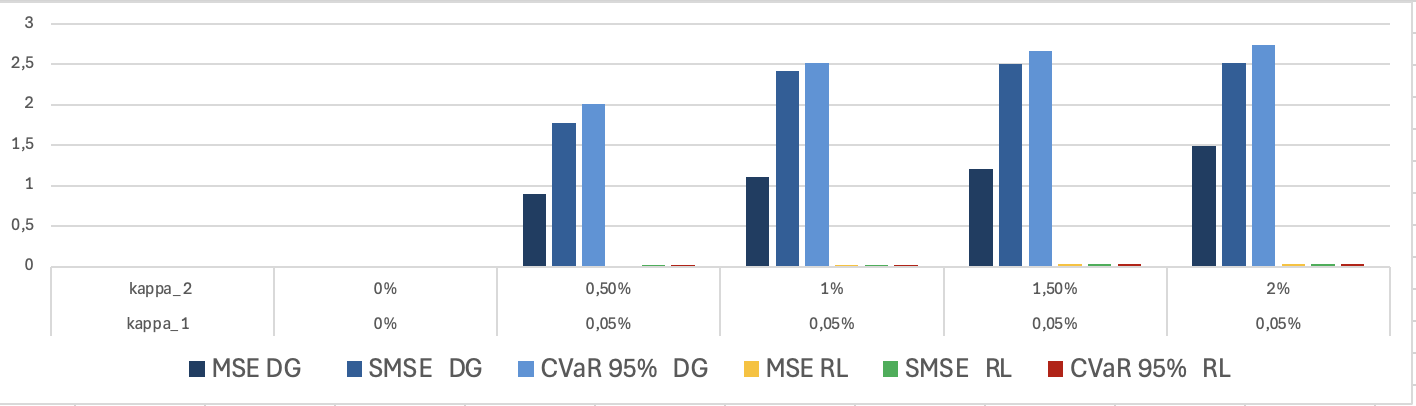}
    \begin{tablenotes}
    \item \small Optimal values are computed across different transaction cost levels using 100,000 out-of-sample paths. The hedged position is an ATM straddle with a maturity of $T = 63$ days. The hedging instrument is an ATM call option with a maturity of $T^{*} = 84$ days.
    \end{tablenotes}
    \label{fig:optimal_threshold_values}
    \end{figure}
\autoref{fig:optimal_threshold_values} reports the optimal rebalancing thresholds $\ell$ across different transaction cost levels for both DG (gray) and RL (blue) strategies, considering all risk measures. Remarkably, the RL algorithms barely rely on the no-trade region, as the optimal values of $\ell$ are close to zero. \autoref{fig:hedging_metrics_straddle} provides further evidence of this phenomenon: risk metrics associated with the RL approaches are minimally affected by the presence of the no-trade region. Such no-trade region is primarily introduced to assist the delta-gamma benchmarks, which are not inherently designed to handle transaction costs efficiently. In the absence of transaction costs, the optimal no-trade region parameter $\ell$ collapses to zero.

\section{Details for the MSGD training approach}\label{appen:MSGDtraining}

The MSGD method estimates the objective function $\mathcal{O}(\theta;\, \lambda)$ by using small samples of the hedging error, referred to as batches. 
Let $\mathbb{B}_{j}
    =\left\{ \xi_{T,i}^{\tilde{\phi}_{\theta_{j}}} \right\}_{i=1}^{B_{\mbox{\scriptsize{batch}}}}$ 
be the $j$-th batch simulated with policy parameters $\theta_{j}$. Using a subset from generated paths, it represents a set of hedging errors
\[
    \xi_{T,i}^{\tilde{\phi}_{\theta_{j}}} 
    = \Psi(S^{(j)}_{T,i}) - V_{T,i}^{\tilde{\phi}_{\theta_{j}}} \quad \mbox{for} \quad i \in \{1, \dots, B_{\mbox{\scriptsize{batch}}}\}, \, j \in \{1, \dots, N_{\mbox{\scriptsize{batch}}} \},
\]
where $S^{(j)}_{T,i}$ and $V_{T,i}^{\tilde{\phi}_{\theta_{j}}}$ respectively represent the time-$T$ underlying asset price and the terminal value of the hedging portfolio for path $i$ of batch $j$. The batch size is $B_{\mbox{\scriptsize{batch}}}=1000$, and the total number of batches is $N_{\mbox{\scriptsize{batch}}}=400$.
The objective function estimates for batch $\mathbb{B}_j$ are
\begin{align*}
    \Hat{\mathcal{O}}^{\left(\mbox{\scriptsize{MSE}}\right)}(\theta_{j}; \lambda,\mathbb{B}_{j}) 
    &= \frac{1}{B_{\mbox{\scriptsize{batch}}}}\sum_{i=1}^{B_{\mbox{\scriptsize{batch}}}} \left(\xi_{T,i}^{\tilde{\phi}_{\theta_{j}}}\right)^2 + \lambda \cdot \widehat{SC}(\theta_{j}, \mathbb{B}_{j}), \\
    \Hat{\mathcal{O}}^{\left(\mbox{\scriptsize{SMSE}}\right)}(\theta_{j}; \lambda,\mathbb{B}_{j}) 
    &= \frac{1}{B_{\mbox{\scriptsize{batch}}}}\sum_{i=1}^{B_{\mbox{\scriptsize{batch}}}} \left(\xi_{T,i}^{\tilde{\phi}_{\theta_{j}}}\right)^2 \mathbbm{1}_{\left\{\xi_{T,i}^{\tilde{\phi}_{\theta_{j}}} \geq 0 \right\}} + \lambda \cdot \widehat{SC}(\theta_{j}, \mathbb{B}_{j}), \\
    \Hat{\mathcal{O}}^{\left(\mbox{\scriptsize{CVaR}}\right)}(\theta_{j};\lambda,\mathbb{B}_{j}) &= \widehat{\mbox{VaR}}_{\alpha}(\mathbb{B}_{j}) + \frac{1}{(1 - \alpha) B_{\mbox{\scriptsize{batch}}}} \sum_{i=1}^{B_{\mbox{\scriptsize{batch}}}} \max\left( \xi_{T,i}^{\tilde{\phi}_{\theta_{j}}} - \widehat{\mbox{VaR}}_{\alpha}(\mathbb{B}_{j}), 0 \right) %\\
    %& \quad 
    + \lambda \cdot \widehat{SC}(\theta_{j}, \mathbb{B}_{j}),
\end{align*}
where
\[
    \widehat{SC}(\theta_{j}, \mathbb{B}_{j}) = \frac{1}{B_{\mbox{\scriptsize{batch}}}} \sum_{i=1}^{B_{\mbox{\scriptsize{batch}}}} \mathbbm{1}_{ \left\{ \max_{t \in \{0, \dots, T\}} \left[P_{t,i} - V_{t,i}^{\tilde{\phi}_{\theta_{j}}} \right] > V_{0,i}^{\tilde{\phi}_{\theta_{j}}} \right\}},
\]
and $\widehat{\mbox{VaR}}_{\alpha}(\mathbb{B}_{j}) = \xi_{T, \left[\lceil \alpha \cdot B_{\mbox{\scriptsize{batch}}} \rceil \right]}^{\tilde{\phi}_{\theta_{j}}}$ is the value-at-risk estimation derived from the ordered sample $\left\{ \xi_{T,[i]}^{\tilde{\phi}_{\theta_{j}}} \right\}_{i=1}^{B_{\mbox{\scriptsize{batch}}}}$, where $\lceil \cdot \rceil$ is the ceiling function. These empirical approximations are used to estimate the gradient of the objective function required in Equation \eqref{updatingrule_1}. The gradient of these empirical objective functions has analytical expressions for FFNN, LSTM and RNN-FNN networks, which can be computed through backpropagation, see for instance \cite{goodfellow2016deep}.

\section{Joint implied volatility and return model}\label{appen:JIVR}

\subsection{Daily implied volatility surface}\label{subappen:IVsurface}

The full functional representation of the IV surface model introduced by \cite{Frnacois2022} is given by:
\begin{align}
     & \sigma (M_{t},\tau_{t},\beta_{t}) =  \underbrace{\beta_{t,1}}_{f_{1}\mbox{: \scriptsize{Long-term ATM IV}}} +  \beta_{t,2}\underbrace{e^{-\sqrt{\tau_{t} /T_{conv}}}}_{f_{2}\mbox{: \scriptsize{Time-to-maturity slope}}} + \beta_{t,3}\underbrace{\left( M_{t}\mathbbm{1}_{\{ M_{t}\geq 0 \}} + \frac{e^{2M_{t}}-1}{e^{2M_{t}}+1}\mathbbm{1}_{\{M_{t}< 0\}} \right)}_{f_{3}\mbox{: \scriptsize{Moneyness slope}}} \nonumber\\
     & + \beta_{t,4}\underbrace{\left( 1-e^{-M_{t}^{2}} \right)\mbox{log}(\tau_{t}/T_{max})}_{f_{4}\mbox{: \scriptsize{Smile attenuation}}}+\beta_{t,5}\underbrace{\left( 1-e^{(3M_{t})^{3}} \right)\mbox{log}(\tau_{t}/T_{max})\mathbbm{1}_{\{M_{t}< 0\}}}_{f_{5}\mbox{: \scriptsize{Smirk}}}\, , \quad \tau_{t} \in [T_{min},T_{max}]. 
     \label{full_5factormodel}
\end{align}

As in \cite{Frnacois2022}, we set $T_{max}=5$ years, $T_{min}=6/252$ and $T_{conv}=0.25$.%\footnote{The moneyness is defined as in \cite{Frnacois2023}: $$M_{t}=M(\tau_{t},S_{t},K,r_{t},q_{t})=\frac{1}{\sqrt{\tau_{t}}}\mbox{log}\frac{S_{t}e^{(r_{t}-q_{t})\tau_{t}}}{K}.$$}

\subsection{Joint implied volatility and return dynamics}\label{subappen:JIVR_timeseries}

The multivariate time series representation of the JIVR model, as introduced by \cite{Frnacois2023}, consists of two key components: one capturing the returns of the underlying asset and another modeling the fluctuations of the implied volatility (IV) surface coefficients. The first component is inspired from the NGARCH(1,1) process with normal inverse Gaussian (NIG) innovations and is formulated as
\begin{align*}
    R_{t+1} &= \xi_{t+1} - \psi(\sqrt{h_{t+1,R}\Delta}) + \sqrt{h_{t+1,R}\Delta}\epsilon_{t+1,R}, \nonumber \\
    h_{t+1,R} &= Y_t + \kappa_R(h_{t,R} - Y_t) + a_R h_{t,R}(\epsilon_{t,R}^2 - 1 - 2\gamma_R \epsilon_{t,R}), \\
    Y_t &= \left( \omega_R \, \sigma \left(0, \frac{1}{12}, \beta_t \right) \right)^2, \nonumber
\end{align*}
where the equity risk premium is
\begin{equation*}
    \xi_{t+1} = \psi(-\lambda \sqrt{h_{t+1,R}\Delta}) - \psi((1-\lambda) \sqrt{h_{t+1,R}\Delta}) + \psi(\sqrt{h_{t+1,R}\Delta}).
\end{equation*}
The innovation process $\{\epsilon_{t,R}\}^T_{t=0}$ is a sequence of iid standardized NIG random variables\footnote{A complete description of the NIG specification is available in \cite{Frnacois2023}.} 
and $\psi$ represents its cumulant generating function.

The evolution of the long-term factor \(\beta_1\) is modeled as
\begin{align*}
    \beta_{t+1,1} &= \alpha_1 + \sum_{i=1}^{5} \theta_{1,j} \beta_{t,j} + \sqrt{h_{t+1,1}\Delta} \epsilon_{t+1,1}, \nonumber \\
    h_{t+1,1} &= U_t + \kappa_1(h_{t,1} - U_t) + a_1 h_{t,1}(\epsilon_{t,1}^2 - 1 - 2\gamma_1 \epsilon_{t,1}), \\
    U_t &= \left( \omega_1  \, \sigma \!\left( 0, \frac{1}{12}, \beta_t \right) \right)^2. \nonumber
\end{align*}

The evolution of the other four IV coefficients, namely for \(i \in \{2, 3, 4, 5\}\), is
\begin{align*}
    \beta_{t+1,i} &= \alpha_i + \sum_{j=1}^{5} \theta_{i,j} \beta_{t,j} + \nu \beta_{t-1,2} \mathds{1}_{\{i=2\}} + \sqrt{h_{t+1,i}\Delta} \epsilon_{t+1,i}, \nonumber \\
    h_{t+1,i} &= \sigma_i^2 + \kappa_i(h_{t,i} - \sigma_i^2) + a_i h_{t,i}(\epsilon_{t,i}^2 - 1 - 2\gamma_i \epsilon_{t,i}),
\end{align*}
where \(\{\epsilon_{t,i} \}_{i=1}^{5}\) are time-independent standardized NIG random variables with parameters \(\{ (\zeta_i, \varphi_i) \}_{i=1}^{5}\). 

The JIVR model imposes a dependence structure on the contemporaneous innovations, i.e., \(\epsilon_t = (\epsilon_{t,R}, \epsilon_{t,1}, ..., \epsilon_{t,5})\), through a Gaussian copula, which is parameterized using a covariance matrix \(\Sigma\) of dimension \(6 \times 6\). Parameter estimates for the entire JIVR model are sourced from Table 5 and Table 6 of \cite{Frnacois2023}.

\section{Benchmarks}\label{appen:benchmarks}

The benchmarks presented in this appendix assume that implied volatilities adhere to the IV model specified in Equation \eqref{5factormodel_linear}.

\subsection{Leland model}\label{appen:Leland_model}

The Leland delta hedging strategy, introduced by \cite{Leland}, modifies the classical option replication framework of \cite{BlackScholes} by incorporating transaction costs, represented by the proportion $\kappa$, and the rebalancing frequency $\lambda$. The hedging position in the underlying asset is given by
\begin{equation*}
   \phi^{(S)}_{t+1} = \mbox{e}^{-q_{t}\tau_{t}}\Phi\left(\tilde{d}_{t} \right),
\end{equation*}
where
\begin{equation*}
\tilde{d}_{t} = \frac{\mbox{log}\left( \frac{S_{t}}{K} \right)+\left( r_{t}-q_{t}+\frac{1}{2}\tilde{\sigma}_{t}^2 \right)\tau_{t}}{\tilde{\sigma}_{t} \sqrt{\tau_{t}}}
\end{equation*}
with the adjusted volatility
\begin{equation*}
\tilde{\sigma}_{t} = \sigma (M_{t},\tau_{t},\beta_{t}) \sqrt{ 1+\sqrt{\frac{2}{\pi}} \frac{2\kappa}{\sigma (M_{t},\tau_{t},\beta_{t})\sqrt{\lambda}} }.
\end{equation*}
Here, $\Phi$ denotes the cumulative distribution function of the standard normal distribution.

\subsection{Delta-gamma hedging}\label{appen:delta_gamma}

The delta-gamma hedging strategy involves both the underlying asset $S$ and an additional hedging instrument, $\text{O}$. This setup allows for neutralizing both the delta and gamma of the portfolio. The trading strategy $\phi$ is fully determined by the process $(\phi^{(S)}, \phi^{(\text{O})})$, expressed as
\begin{equation*}
    (\phi^{(S)}_{t+1}, \phi^{(\text{O})}_{t+1}) 
    = \left(\Delta^{\mathcal{P}}_{t} - \frac{\Gamma^{\mathcal{P}}_{t}}{\Gamma_{t}^{(\text{O})}} \Delta^{(\text{O})}_{t}, \, \frac{\Gamma^{\mathcal{P}}_{t}}{\Gamma_{t}^{(\text{O})}}\right),
\end{equation*}
where $\Delta^{\mathcal{P}}_{t}$, $\Gamma^{\mathcal{P}}_{t}$, and  $\Delta^{(\text{O})}_{t}$, $\Gamma_{t}^{(\text{O})}$ represent the delta and gamma of the hedged portfolio and of the hedging option, respectively. The self-financing constraint \eqref{eq:self-financing} fully determines $\phi^{(r)}_{t+1}$.
For all Greeks we use the implied volatility $\sigma (M_{t},\tau_{t},\beta_{t})$ from the static surface as the volatility input parameter.

\subsection{No-trade region}\label{appen:delta_gamma no trade region}
This is a recursive construction.  In the time interval $(t-1,t]$, denote the hedging portfolio with the no-trade threshold $\ell$ by $\phi_{t}^{(\ell)} = \left(\phi^{(\ell,r)}_{t}, \phi^{(\ell,S)}_{t}, \phi^{(\ell,\text{O})}_{t}\right)$. At time $t$, its value is  
\begin{equation*}
    V^{(\ell,\phi)}_{t}=\phi^{(\ell,r)}_{t} \text{e}^{r_{t} \Delta} + \phi^{(\ell,S)}_{t} S_{t} \text{e}^{q_{t} \Delta} + \phi^{(\ell,\text{O})}_{t} O_{t}\left( T^*\right).
\end{equation*}
The no-trade region constraint is set up such that
\begin{align*}
    \left(\phi^{(\ell,S)}_{t+1}, \phi^{(\ell,\text{O})}_{t+1}\right) &= 
    \begin{cases} 
        \left(\phi^{(\ell,S)}_{t}, \phi^{(\ell,\text{O})}_{t}\right) , & \text{if } \left\vert \phi_{t}^{(\ell, S)} - \left( \Delta^{\mathcal{P}}_{t} - \frac{\Gamma^{\mathcal{P}}_{t}}{\Gamma_{t}^{(\text{O})}} \Delta^{(\text{O})}_{t}\right) \right\vert + \left\vert \phi^{(\ell, \text{O})}_{t} - \frac{\Gamma^{\mathcal{P}}_{t}}{\Gamma_{t}^{(\text{O})}} \right\vert \leq \ell, \\[4pt]
        \left(\Delta^{\mathcal{P}}_{t} - \frac{\Gamma^{\mathcal{P}}_{t}}{\Gamma_{t}^{(\text{O})}} \Delta^{(\text{O})}_{t}, \, \frac{\Gamma^{\mathcal{P}}_{t}}{\Gamma_{t}^{(\text{O})}}\right) , & \text{otherwise}.
    \end{cases}
\end{align*}
The bank account position is $$\phi^{(\ell,r)}_{t+1}=V^{(\ell,\phi)}_{t} - \phi^{(\ell,S)}_{t+1} S_{t} - \phi^{(\ell,O)}_{t+1} O_{t}(T^*) -\kappa_1 \left\vert \phi^{(\ell,S)}_{t+1}-\phi^{(\ell,S)}_{t}\right\vert S_{t} -\kappa_2 \left\vert \phi^{(\ell,O)}_{t+1}-\phi^{(\ell,O)}_{t}\right\vert O_{t}(T^*).$$
The parameter $\ell$ is optimized by minimizing one of the three objective functions computed on the entire learning set:
\begin{align*}  
    \Hat{\mathcal{O}}^{\left(\mbox{\scriptsize{MSE}}\right)}(\ell) 
    &= \frac{1}{N}\sum_{i=1}^N \left(\xi_{T,i}^{\phi^{(\ell)}}\right)^2 
    \\    \Hat{\mathcal{O}}^{\left(\mbox{\scriptsize{SMSE}}\right)}(\ell) 
    &= \frac{1}{N}\sum_{i=1}^N \left(\xi_{T,i}^{\phi^{(\ell)}}\right)^2 \mathbbm{1}_{\left\{\xi_{T,i}^{\phi^{(\ell)}} \geq 0 \right\}} 
    \\    \Hat{\mathcal{O}}^{\left(\mbox{\scriptsize{CVaR}}\right)}(\ell) &= \widehat{\mbox{VaR}}_{\alpha} + \frac{1}{(1 - \alpha) N} \sum_{i=1}^N \max\left( \xi_{T,i}^{\phi^{(\ell)}} - \widehat{\mbox{VaR}}_{\alpha}, 0 \right)
\end{align*}
where $\xi_{T,i}^{\phi^{(\ell)}} = \mathcal{P}_{T,i} - V_{T,i}^{\phi^{(\ell)}} $ and $\widehat{\mbox{VaR}}_{\alpha} = \xi_{T, \left[\lceil \alpha \cdot N \rceil \right]}^{\phi^{(\ell)}}$ is the value-at-risk estimate derived from the ordered sample $\left\{ \xi_{T,[i]}^{\phi^{(\ell)}} \right\}_{i=1}^N$.

\section{Impact of state variable inclusion on hedging performance}\label{appen:state_space}

To evaluate the impact of including state variables $\mathcal{P}_{t}$, $\Delta_{t}^{P}$, and $\gamma_{t}^{P}$ in the reinforcement learning framework, we conduct additional numerical experiments. Specifically, we compare the performance of RL agents trained with and without these variables across various risk measures. 
\autoref{table:state_variables_evaluation} demonstrates that the inclusion of state variables consistently improves hedging performance because they provide additional structure, which helps with the training.

\begin{table}[H]
    \centering
    \caption{Optimal risk measure values for different state space configurations.}
    \label{table:state_variables_evaluation}
    \renewcommand{\arraystretch}{1.5}
    %\resizebox{\textwidth}{!}{
    \begin{tabular}{{p{2.5cm}>{\centering\arraybackslash}p{0.01cm}>{\centering\arraybackslash}p{1.5cm}>{\centering\arraybackslash}p{0.01cm}>{\centering\arraybackslash}p{1.5cm}>{\centering\arraybackslash}p{0.01cm}>{\centering\arraybackslash}p{1.5cm}}}
    \toprule
    \multicolumn{1}{l}{State space} & & \multicolumn{1}{c}{MSE} & & \multicolumn{1}{c}{SMSE} & & \multicolumn{1}{c}{CVaR$_{95\%}$}\\
    \midrule
    $\mathcal{S} \backslash \{\mathcal{P}_{t}, \Delta_{t}^{P}, \gamma_{t}^{P} \}$ & & 0.195 & & 0.089 & & 0.696 \\
    $\mathcal{S} \backslash \{\mathcal{P}_{t}\}$ & & 0.128 & & 0.069 & & 0.680\\
    $\mathcal{S}$ & & {\bfseries 0.094} & &	{\bfseries 0.022} & & {\bfseries 0.502}\\
    \bottomrule
    \end{tabular}%}
    \begin{tablenotes}
    \item \small Optimal values are computed using 400,000 during training. Transaction cost levels are set to $\kappa_{1}=\kappa_{2}=0\%$. The hedge consists of an ATM straddle with a maturity of $T = 63$ days and an average value of \$7.55. The hedging instrument is an ATM call option with a maturity of $T^{*} = 84$ days. The full state space, as described in \autoref{tab:state_variables}, is denoted by $\mathcal{S}$.
    \end{tablenotes}
\end{table}

\section{Statistical arbitrage}\label{appen:statistical_arbitarge}

This analysis examines whether our framework can embed a speculative layer, such as statistical arbitrage, by leveraging the structural properties of the risk measure that guides the hedging optimization process.

Following the definition in \cite{assa2013hedging} and studies such as \cite{buehler2021deep}, \cite{horikawa2024relationship}, and \cite{François2025_stat_arbitrage}, we define statistical arbitrage strategies as profit-seeking trading strategies that exploit the blind spots of the risk measure.

Specifically, we assess whether the difference between RL strategies, $\phi^{RL}$, and DG strategies, $\phi^{DG}$, denoted as  
\begin{equation*}
    \phi^{-} = \phi^{RL} - \phi^{DG},
\end{equation*}
exhibits statistical arbitrage characteristics with respect to a risk measure $\rho$. More precisely, we examine whether  
\begin{equation*}
    \rho\left(-V_{T}^{\phi^{-}}(0)\right) < 0
\end{equation*}  
occurs. This condition implies that the strategy that requires no initial investment is strictly less risky than a null investment according to $\rho$. We investigate whether $\phi^{-}$ behaves as statistical arbitrage within our framework, analyzing whether RL merely introduces a speculative component to the DG strategy or if another mechanism is at play. This analysis is conducted using CVaR$_{95\%}$ and SMSE as risk measures.

\autoref{table:statistical_arbitrage_stats} presents the hedging error risk associated with the trading strategy $\phi^{-}$, which represents the differential position between the RL and DG strategies. This analysis is conducted across the strategies obtained under different risk measures while hedging an ATM straddle intrument with a maturity of $T=63$ days. 

\begin{table}[h]
    \centering
    \renewcommand{\arraystretch}{1.5}
    \caption{Statistical arbitrage statistic.}
    \label{table:statistical_arbitrage_stats}
    \begin{tabular}{{p{1.5cm}>{\centering\arraybackslash}p{0.1cm}>{\centering\arraybackslash}p{2.5cm}>{\centering\arraybackslash}p{2.5cm}>{\centering\arraybackslash}p{2.5cm}>{\centering\arraybackslash}p{2.5cm}>{\centering\arraybackslash}p{2.5cm}}}
    \toprule
     \multicolumn{2}{c}{} & \multicolumn{5}{c}{$\rho\left(-V_{T}^{\phi^{-}}(0)\right)$} 
    \smallskip \\
    \cline{3-7}
    Risk measure & & $\kappa_{1}=\kappa_{2}=0\%$ & $\kappa_{2}$=0.5\% & $\kappa_{2}$=1\% & $\kappa_{2}$=1.5\% & $\kappa_{2}$=2\% \\
    \midrule
    %MSE &  & 1.809 & 1.452 & 1.466 & 1.524 & 1.619 \\ 
    SMSE &  & 1.719 & 1.597 & 1.691 & 1.805 & 1.882 \\ 
    CVaR$_{95\%}$ &  & 1.721 & 1.583 & 1.644 & 1.782 & 1.767 \\ 
    \bottomrule
    \end{tabular}
    \begin{tablenotes}
    \item \small Results are computed over 100,000 out-of-sample paths according to the conditions outlined in \cref{subsub:network_architecture}. The hedge consists of an ATM straddle with a maturity of $T = 63$ days. The hedging instrument is an ATM call option with a maturity of $T^{*} = 84$ days. The transaction cost for the underlying asset is set to $\kappa_{1}=0.05\%$, except for the first column where $\kappa_{1}=0\%$.
    \end{tablenotes}
\end{table}

\begin{figure}[H]\centering
    \caption{P\&L distribution for the strategy $\phi^{-}$.}
    \includegraphics[width=16.5cm]{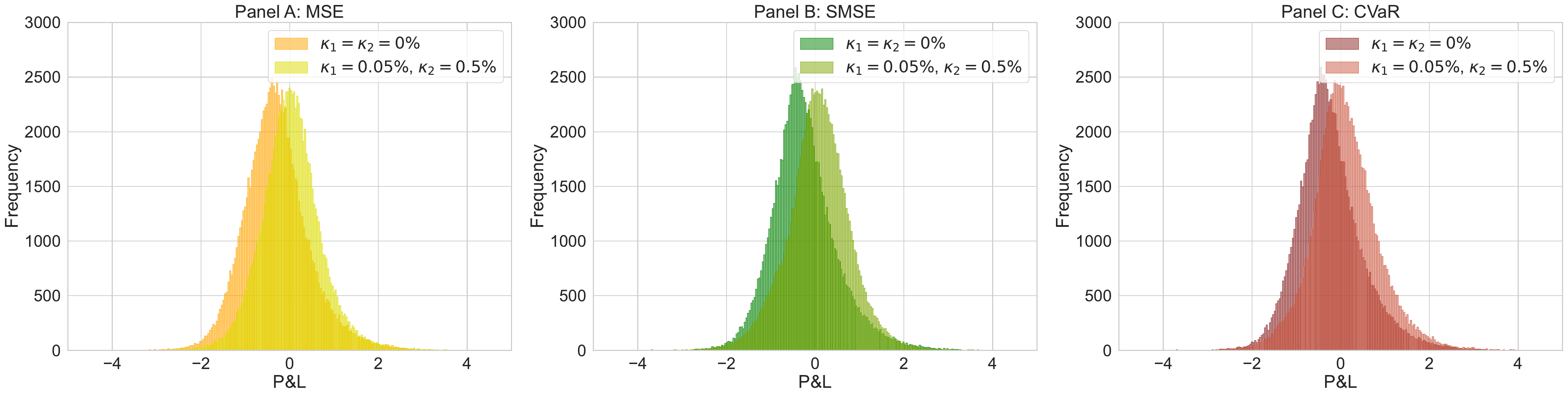}
    \begin{tablenotes}
    \item \small Distributions are computed using 100,000 out-of-sample paths. The P\&L is simply defined by the portfolio value $V_{T}^{\phi^{-}}(0)$ at maturity. The hedge consists of an ATM straddle with a maturity of $T = 63$ days. The hedging instrument is an ATM call option with a maturity of $T^{*} = 84$ days.
    \end{tablenotes}
    \label{fig:statistical_arbitrage}
\end{figure}

Our numerical results show no evidence of statistical arbitrage, as all hedging error risks produce positive values. To further illustrate the absence of arbitrage-like behavior, \autoref{fig:statistical_arbitrage} presents the profit and losses (P\&L) of the strategy $\phi^{-}$ at time $T$ with no initial investment, considering two scenarios: one without transaction costs and another with transaction cost levels set at 0.05\% for $\kappa_{1}$ and 0.5\% for $\kappa_{2}$. The three panels display distributions that are either symmetric around zero or shifted to the left, indicating the absence of profit-seeking trading strategies. This reinforces the conclusion that the RL strategies within our framework are solely focused on hedging, without introducing speculative overlays.

\pagebreak
\pagebreak
\centerline{\Large \textbf{Supplementary material (not part of the paper)}}

\section{Systematic outperformance of RL agents}\label{appen:systematic_outperformance}

\begin{figure}[H]\centering
    \caption{Empirical distribution of risk measures.}
    \includegraphics[width=16cm]{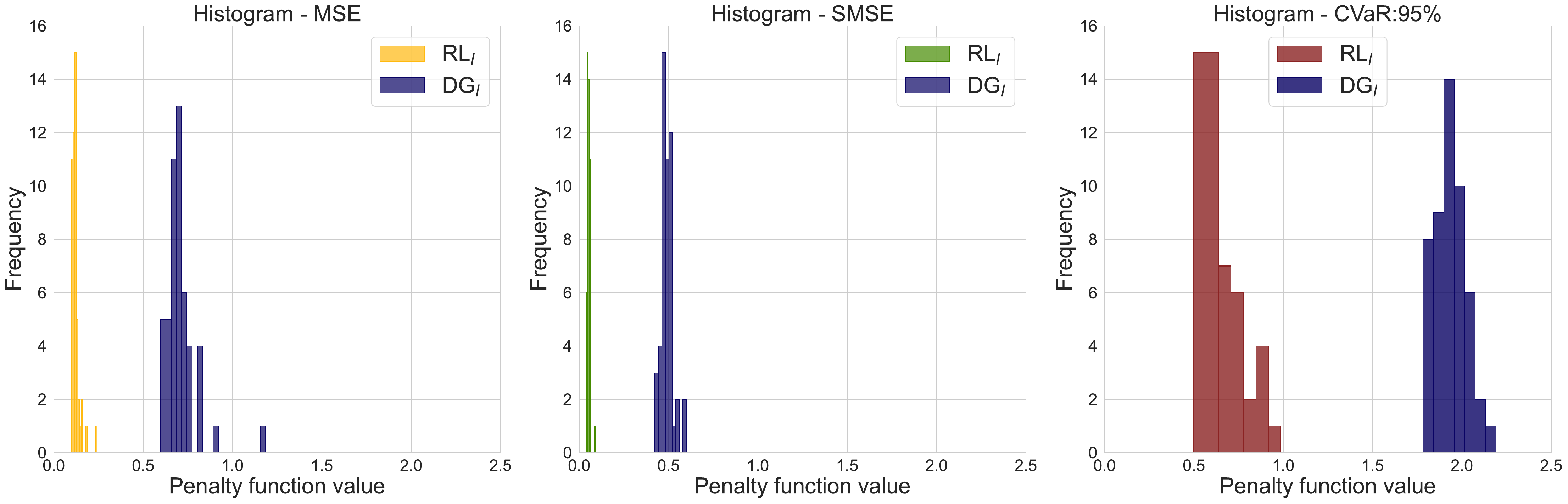}
    \begin{tablenotes}
    \item \small Results are computed using bootstrapping with a sample size of 1,000 over 100,000 out-of-sample paths according to the conditions outlined in \cref{subsub:network_architecture}. The hedge consists of an ATM straddle with a maturity of $T = 63$ days and an average value of \$7.55. The hedging instrument is an ATM call option with a maturity of $T^{*} = 84$ days. Transaction cost levels are set to 0.05\% for $\kappa_1$ and 0.5\% for $\kappa_2$.
    \end{tablenotes}
    \label{fig:systematic_outperformance}
\end{figure}

We validate the outperformance of RL agents by hedging a straddle instrument with a maturity of $T = 63$ days, incorporating an ATM call option with a maturity of $T^{*} = 84$ days as a hedging instrument. In this validation, we analyze the empirical distribution of each risk measure under transaction cost levels set to $\kappa_{1} = 0.05\%$ and $\kappa_{2} = 0.5\%$ for simplicity. The empirical distributions are derived by bootstrapping the hedging error over 100,000 paths, with batches of size 1,000. As shown in \autoref{fig:systematic_outperformance}, the RL approach consistently outperforms the delta gamma strategy, as evidenced by the non-overlapping empirical distributions.

\section{JIVR Model parameters}
%------------------------------
%with parameters \(\zeta_R\) and \(\varphi_R\). 

The standardized NIG random variable \(\epsilon\) has the two-parameter NIG density function 
$$f(x)=\frac{B_{1}\left( \sqrt{\frac{\varphi^{6}}{\varphi^{2}+\zeta^{2}}+(\varphi^{2}+\zeta^{2})\left(x+ \frac{\varphi^{2}\zeta}{\varphi^{2}+\zeta^{2}} \right)^{2}} \right)}{\pi\sqrt{\frac{1}{\varphi^{2}+\zeta^{2}}+\frac{\varphi^{2}+\zeta^{2}}{\varphi^{6}}\left(x+ \frac{\varphi^{2}\zeta}{\varphi^{2}+\zeta^{2}} \right)^{2}} } e^{\left( \frac{\varphi^{4}}{\varphi^{2}+\zeta^{2}}+\zeta\left(x+ \frac{\varphi^{2}\zeta}{\varphi^{2}+\zeta^{2}} \right) \right),}$$ where $B_{1}(\cdot)$ denotes the modified Bessel function of the second kind with index 1. 
The standard four-parameter $(\alpha, \beta, \delta, \mu)$ density function can be recovered by setting $\beta = \zeta$ and $\sqrt{\alpha^2 - \beta^2} = \varphi$, while enforcing a zero mean and unit variance to express $\delta$ and $\mu$ in terms of $\alpha$ and $\beta$. The parameters governing the excess return component of the model are given by 
$$(\Theta_R = (\lambda, \kappa_R, \gamma_R, a_R, \omega_R, \zeta_R, \varphi_R).$$ 
Parameters for the IV coefficient marginal processes are denoted
$$\{ \Theta_i = (\omega_1, \alpha_i, \theta_{i,1}, \theta_{i,2}, \theta_{i,3}, \theta_{i,4}, \theta_{i,5}, \nu, \sigma_i, \kappa_i, a_i, \gamma_i, \zeta_i, \varphi_i)\}_{i=1}^{5}.$$

\begin{table}[H]
\centering
\renewcommand{\arraystretch}{1.5}
\caption{Estimated Gaussian copula parameters.}
\begin{tabular}{lrrrrrr}
\toprule
 & $\epsilon_{t,R}$ & $\epsilon_{t,1}$ & $\epsilon_{t,2}$ & $\epsilon_{t,3}$ & $\epsilon_{t,4}$ & $\epsilon_{t,5}$ \\
\midrule
$\epsilon_{t,R}$ & 1.000 &  &  &  & &  \\
$\epsilon_{t,1}$ & -0.550 & 1.000 &  &  &  &  \\
$\epsilon_{t,2}$ & -0.690 & 0.140 & 1.000 &  & &  \\
$\epsilon_{t,3}$ & 0.030 & -0.030 & -0.010 & 1.000 &  &  \\
$\epsilon_{t,4}$ & -0.220 & 0.250 & 0.120 & 0.280 & 1.000 &  \\
$\epsilon_{t,5}$ & -0.340 & 0.170 & 0.370 & 0.130 & -0.050 & 1.000 \\
\bottomrule
\end{tabular}
\end{table}

\begin{table}[H]
\centering
\renewcommand{\arraystretch}{1.5}
\caption{ JIVR model parameter estimates. }
\begin{tabular}{lrrrrrrr}
\toprule
Parameter  & $\beta_1$ & $\beta_2$ & $\beta_3$ & $\beta_4$ & $\beta_5$ & & S\&P500 \\
\cline{1-6}\cline{8-8}
$\alpha$ & 0.000899 & 0.008400 & 0.000770 & -0.001393 & 0.000657 & $\lambda$ & 2.711279 \\
$\theta_1$ & 0.996290 & -0.013869 &  & 0.002841 &  &  &\\
$\theta_2$ & 0.003669 & 0.877813 & 0.001300 &  &  &  &\\
$\theta_3$ &  & -0.032640 & 0.997071 & 0.003722 & -0.004198 &  &\\
$\theta_4$ &  &  &  & 0.980269 &  &  &\\
$\theta_5$ &  & -0.047789 &  &  & 0.986019 &  &\\
$\nu$ &  & 0.089445 &  &  &  &  &\\
$\sigma\sqrt{252}$ &  & 0.380279 & 0.052198 & 0.048641 & 0.051536 &  &\\
$\omega$ & 0.267589 &  &  &  &  & & 0.977291 \\
$\kappa$ & 0.838220 & 0.965751 & 0.974251 & 0.945377 & 0.980844 & & 0.888977 \\
$a$ & 0.134152 & 0.098272 & 0.092646 & 0.102201 & 0.100502 & & 0.056087 \\
$\gamma$ & -0.111813 & -1.482862 & 0.096766 & 0.060558 & -0.102996 & & 2.507796 \\
$\zeta$ & 0.143760 & 0.852943 & 0.029109 & -0.159051 & 0.092664 & & -0.641306 \\
$\varphi$ & 1.351070 & 1.538928 & 2.284780 & 1.449977 & 1.428477 & & 2.039669 \\
\bottomrule
\end{tabular}
\end{table}

\section{Impact of no-trade regions}\label{subsub:rebalancing}

Since the no-trade region is determined by the rebalancing threshold, we assess its impact by examining how it influences both the rebalancing frequency and hedging cost. The rebalancing frequency, defined as the proportion of days on which portfolio positions are adjusted along a given path, is given by  
\begin{equation}
    \mbox{RF}_{l} = \frac{1}{T}\sum_{t=0}^{T-1} \mathbbm{1}_{\left\{\phi_{t+1}\neq \phi_{t}\right\}}.
\end{equation}
The hedging cost  
\begin{equation}
    \mbox{HC}_{l} 
    = \sum_{t=0}^{T-1} e^{-r\Delta t} \mathcal{HC}_{t},
\end{equation}
is the sum of discounted transaction costs over a given path where the transaction cost at time $t$, $\mathcal{HC}_{t}$, is 
\begin{equation}
    \mathcal{HC}_{t} = \kappa_{1} S_{t} \mid \phi_{t+1}^{(S)} - \phi_{t}^{(S)} \mid + \kappa_{2} \text{O}_{t}(T^{*}) \mid \phi^{(\text{O})}_{t+1} - \phi^{(\text{O})}_{t} \mid.
\end{equation}

\begin{figure}[H]\centering
    \caption{Rebalancing frequency and average hedging transaction costs.}
    \includegraphics[width=17cm]{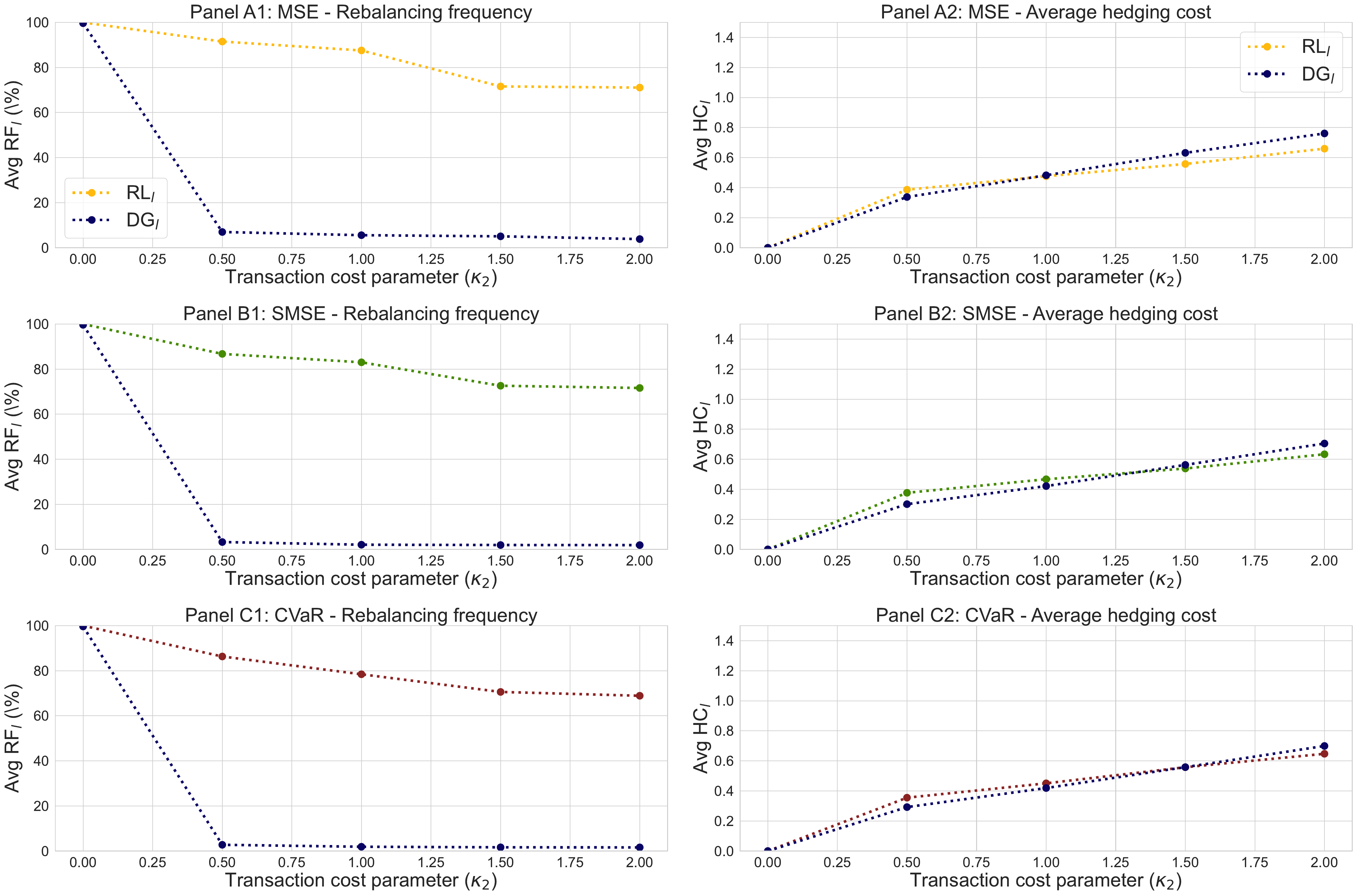}
    \begin{tablenotes}
    \item \small Results are computed over 100,000 out-of-sample paths according to the conditions outlined in \cref{subse:fine-tuning}.
    \end{tablenotes}
    \label{fig:rebalancing}
\end{figure}

This analysis evaluates the trade-off between portfolio adjustment frequency and transaction costs. \autoref{fig:rebalancing} illustrates the effect of the transaction costs on both rebalancing frequency and hedging cost across all risk measures and transaction cost levels.

Results depicted in \autoref{fig:rebalancing} show that RL agents resort to a higher average rebalancing frequency compared to DG strategies, which tend to behave more like semi-static approaches with fewer rebalancing days. This finding aligns with the observations of \cite{carr2014static}, who show that increasing the rebalancing frequency does not necessarily improve the performance of option tracking frameworks such as delta hedging in the presence of transaction cost. 

Conversely, as $\kappa_2$ increases, RL agents retain high rebalancing frequency, but keep average transaction costs to a level similar to DG. Thus, more gradual and frequent adjustments from RL mitigate risk more effectively than DG as documented in \cref{subsub:benchmarking_tc}, while leading to similar transaction costs.

%%%%%%%%%%%%%%%%%%%%%%%%

\section{Soft constraint regularization}\label{appen:soft_constraint}

The estimation of the penalization parameter 
$\lambda$ introduced in Equation \eqref{eq:penaltyfunction}, which governs the weight of the soft constraint in the optimization process, is approached as a model selection problem. In this framework, the model is trained multiple times using fixed values of $\lambda$, iterating across four different values for $\lambda$. 

\begin{figure}[H]\centering
    \caption{Risk measure and soft constraint values.}
    \includegraphics[width=17cm]{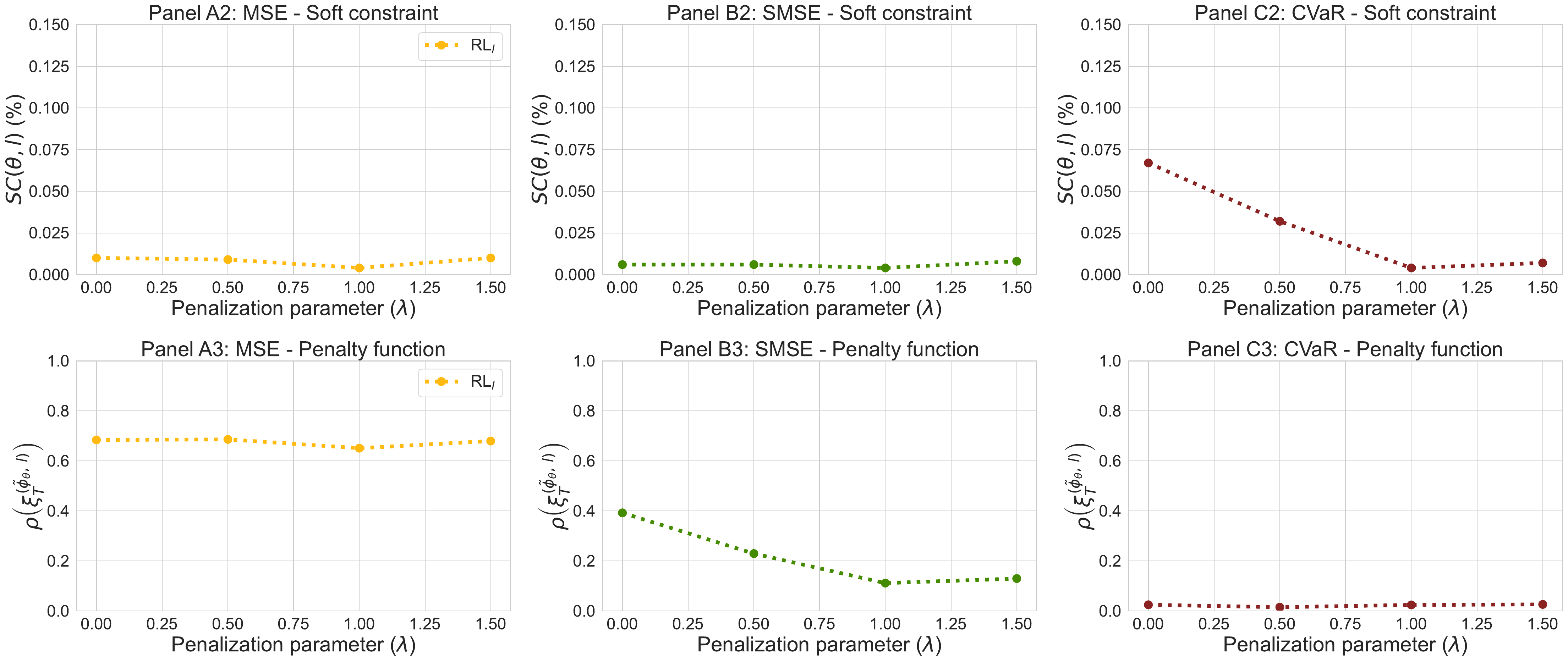}
    \begin{tablenotes}
    \item \small Results are computed over 100,000 out-of-sample paths according to the conditions outlined in \cref{subsub:network_architecture}. The hedge consists of an ATM straddle with a maturity of $T = 63$ days and an average value of \$7.55. The hedging instrument is an ATM call option with a maturity of $T^{*} = 84$ days.
    \end{tablenotes}
    \label{fig:soft_constraint}
\end{figure}

The optimal $\lambda$ is then selected based on an evaluation conducted on the validation set,\footnote{The validation set consists of 100,000 independent simulated paths, generated as outlined in \cref{subsec:market_parameters}. This set is distinct from the training and test sets described in \cref{subsub:network_architecture}.} considering two key factors: the soft constraint value and the risk measure. To determine the optimal $\lambda$, we hedge an ATM straddle with a maturity of $T = 63$ days, assuming no transaction costs ($\kappa_{1} = \kappa_{2} = 0\%$). The hedging strategy optimization considers three risk measures: MSE, SMSE, and $\text{CVaR}_{95\%}$. This process is repeated for different values of $\lambda$: 0, 0.5, 1, and 1.5. \autoref{fig:soft_constraint} presents the optimal soft constraint values and risk measure outcomes for each $\lambda$, evaluated on a validation set.

The results illustrated in \autoref{fig:soft_constraint} highlight the heightened sensitivity to variations in the penalization parameter $\lambda$ when using asymmetric risk measures. The SMSE risk measure exhibits significant sensitivity of $\rho$, achieving its minimum value at $\lambda = 1$, which aligns with the corresponding minimum value of the soft constraint penalty. For the CVaR, the soft constraint penalty demonstrates greater sensitivity compared to the risk measure itself, indicating that CVaR is more susceptible to higher tracking error in the absence of the soft constraint.

The minimum value of the soft constraint penalty for CVaR also occurs at $\lambda = 1$, corresponding to the stabilization point of the risk measure. In contrast, the MSE risk measure is mildly affected by the soft constraint. Yet its minimum value is also observed at $\lambda = 1$, mirroring the behavior of the other risk measures.

Based on these findings, we select $\lambda = 1$ for our subsequent experiments. This value leads to soft constraint penalty levels that remain below $0.025\%$ across all risk measures, minimizing the likelihood of observing paths with large tracking error.

%%%%%%%%%%%%%%%%%%%

\section{In-sample backtest}

In this section, we benchmark our approach using historical paths generated by the JIVR model, covering the period from January 5, 1996, to December 31, 2020, to assess the effectiveness of RL agents. This experiment evaluates the performance of risk management strategies based on the historical series $(R_{t}, \beta_{t})$.  
Hedging performance is assessed by introducing a new ATM straddle instrument with a 63-day maturity every 21 business days along the historical paths. The initial hedging portfolio values are set equal to the straddle prices, which are computed using the prevailing implied volatility surface on the day the hedge is initiated.

\begin{figure}[H]\centering
    \caption{Cumulative P\&L for the hedge of ATM straddles under real asset price dynamics.}
    \includegraphics[width=17.5cm]{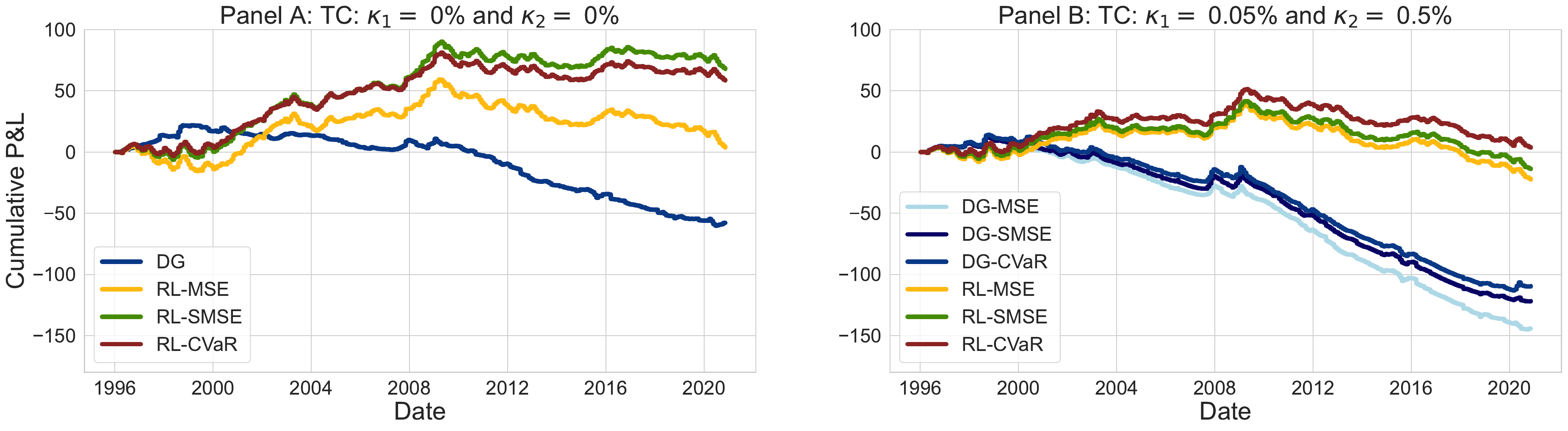}
    \begin{tablenotes}
    \item \small Results are computed based on the observed P\&L from hedging 296 straddle positions with maturity 63-days under real market conditions observed from May 1, 1996, to December 31, 2020. A new ATM straddle is considered every 21 business days. Agents are trained according to the conditions outlined in \cref{subse:fine-tuning} using an ATM call option with a maturity of $T^{*} = 84$ days as the hedging instrument.
    \end{tablenotes}
    \label{fig:cumulative_p&l}
\end{figure}

To evaluate the robustness of our approach under diverse market conditions, we compare cumulative P\&Ls. The cumulative P\&L at a given date is defined as the sum of the total P\&L generated by all straddle trades whose hedging period has expired.
\autoref{fig:cumulative_p&l} illustrates the evolution of cumulative P\&Ls, where each of the two panels correspond to different transaction cost levels.

As illustrated in \autoref{fig:cumulative_p&l}, RL strategies consistently outperform the benchmarks in both scenarios, namely with and without transaction costs. Notably, the gap between the cumulative P\&L of RL agents and the benchmarks widens significantly as transaction costs increase, highlighting the adaptability of the RL approach to transaction costs across diverse market conditions. 
Additionally, RL strategies optimized using the MSE function yield lower cumulative P\&L compared to those optimized with asymmetric risk measures, reflecting the inherent differences in the objectives of these risk measures.

\begin{figure}[H]\centering
    \caption{Hedging error distribution for a ATM straddle instrument with a maturity of 63 days under real asset price dynamics.}
    \includegraphics[width=17.5cm]{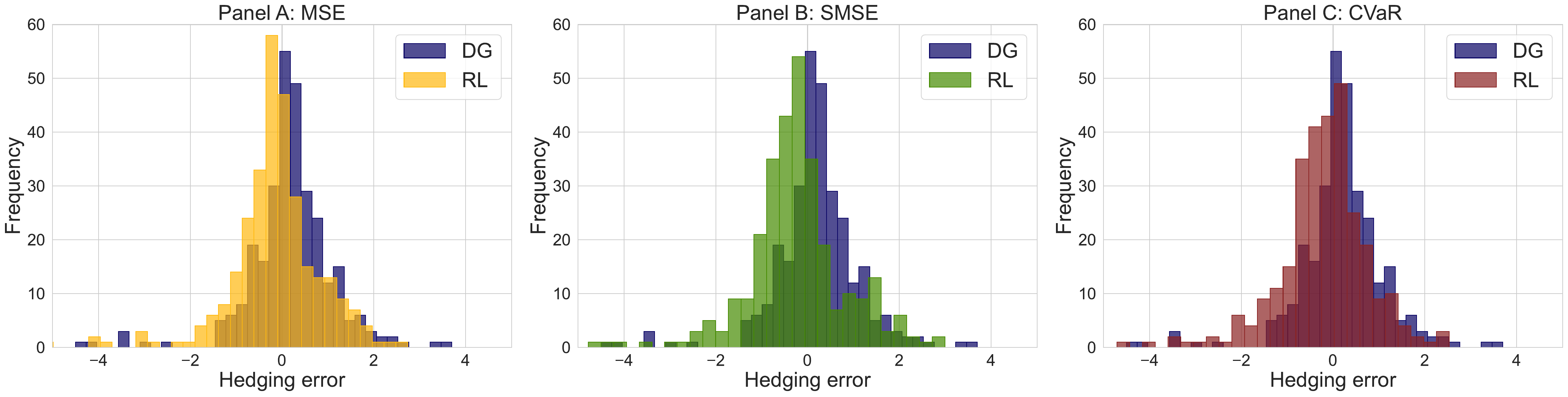}
    \begin{tablenotes}
    \item \small Results are computed based on the observed P\&L from hedging 296 ATM straddle instruments with maturity  of $T = 63$ under real market conditions observed from May 1, 1996, to December 31, 2020. The hedging instrument is an ATM call option with a maturity of $T^{*} = 84$ days. Transaction cost levels are set to 0\%.
    \end{tablenotes}
    \label{fig:hedging_error_backtest}
\end{figure}

To evaluate hedging errors under real asset price dynamics, we analyze the distribution of terminal errors generated by 296 ATM straddles from May 1, 1996, to December 31, 2020. \autoref{fig:hedging_error_backtest} presents the histogram of hedging errors for benchmark strategies and RL agents across all risk measures, without transaction costs.

As shown in \autoref{fig:hedging_error_backtest}, RL strategies exhibit a hedging error distribution that is shifted towards the left, highlighting greater profitability and lower downside risk.
These findings highlight the robustness of the RL approach to different market conditions and transaction cost levels.

\end{document}